\documentclass[prb,twocolumn,showpacs]{revtex4}
\usepackage{graphicx}
\usepackage{dcolumn}
\usepackage{bm}
\usepackage{amsmath}

\begin{document}

\title{Symmetry-projected variational approach for 
ground and excited states of
the two-dimensional Hubbard model}

\author{R. Rodr\'{\i}guez-Guzm\'an$^{1,2}$, K. W. Schmid$^{3}$, Carlos A. Jim\'enez-Hoyos$^{1}$
and Gustavo E. Scuseria$^{1,2}$}

\affiliation{
$^{1}$ Department of Chemistry, Rice University, Houston, Texas 77005, USA
\\
$^{2}$ Department of  Physics and Astronomy, Rice University, Houston, Texas 77005, USA
\\
$^{3}$ Institut f\"ur Theoretische Physik der Universit\"at T\"ubingen, Auf der Morgenstelle 14, D-72076  
T\"ubingen, Germany
}

\date{\today} 

\begin{abstract}
We present a symmetry-projected configuration mixing scheme to describe ground and 
excited states, with well defined quantum numbers, of the two-dimensional Hubbard model with nearest-neighbor 
hopping and periodic boundary conditions. Results 
for the half-filled $2 \times 4$, $4 \times 4$, and $6 \times 6$ lattices, as well as doped $4 \times 4$ systems, compare 
well with available results, both exact and from other state-of-the-art approximations. We report spectral functions 
and density of states obtained from a well-controlled ansatz for the $(N_{e} \pm 1)$-electron system. Symmetry projected methods have 
been widely used for the many-body nuclear physics problem but have received little attention in the solid state community. Given their
 relatively low (mean-field) computational cost and the high quality of results here reported, we believe that they deserve further scrutiny.
\end{abstract}


\maketitle

\section{Introduction}
\label{introd}
Since  the discovery of high-Tc superconductivity, \cite{HTCSC-1} there has been a growing 
interest in  the properties of correlated 
two-dimensional (2D)
electronic systems. \cite{Dagotto-review} Within this context, the 
Hubbard model \cite{Hubbard-model_def1}  has received a lot of attention since it is considered
one of the 
simplest models still  containing 
the relevant physics. \cite{Anderson-suggestion}
  Renewed interest in the 
Hubbard Hamiltonian  also comes from recent 
experiments
\cite{optical-1,optical-2}
 with cold fermionic atoms in optical lattices which open 
the possibility 
for direct simulations of the  model with lattice emulators. \cite{IBloch-revmodphys}
 Hubbard-like models are also relevant to describe electronic
properties within the active research field of graphene. \cite{CastroNeto-review}

The repulsive  Hubbard Hamiltonian 
is a very 
interesting  model in theoretical physics. On the one hand, neither its 
hopping  (one-body) nor its  on-site interaction (two-body) terms
favor any interesting magnetic ordering. On 
the other hand, when both of them combine 
into the full  Hamiltonian  a rich variety of interesting
phenomena 
is displayed, for example, correlation-driven metal-insulator transitions, \cite{Gebhard-1}
 ferromagnetism, \cite{Nagaoka-theorem} 
deviations from the standard Fermi-liquid results, \cite{Dagotto-Sch-nfl} long-wavelength collective modes \cite{Chang-Shiwei}
and spatially inhomogeneous phases. \cite{ChiaChen-Shiwei} The
 dimensionality of the model also 
challenges the theoretical tools at our disposal. Exact 
analytical solutions exist in the one-dimensional (1D) case
\cite{text-Hubbard-1D} whereas the present knowledge 
of the basic  
quantum mechanical properties of the 2D Hubbard  Hamiltonian relies, to a large
extent, on numerical techniques applied to the Hamiltonian itself or to
its strong coupling approximations, i.e., the t-J, t-J$^{*}$ 
and  Heisenberg 
models. \cite{Dagotto-review,Szczepanski,Review-Heisenberg-model} In particular, for the case
of the full 2D Hubbard Hamiltonian, a very efficient Lanczos
algorithm, \cite{Lanczos-Fano} based  on the classification of all the irreducible 
representations of the space group, has allowed 
systematic studies  in the $4 \times 4$ lattice.

Going beyond the present limits of exact diagonalization (ED) techniques 
requires a truncation strategy.  A key issue is then  how to 
truncate the model space while still being able to retain the 
most important degrees of freedom relevant for the description of 
a particular ground and/or 
excited state. Nowadays 
there are several  methods at our disposal, some of them already heavily 
used to study 1D and 2D Hubbard models with variable degree of success.
One that has been used with great 
success is the Quantum Monte Carlo \cite{Nightingale,Raedt-MC,Sorella-1} (QMC) approach. 
Another is the density matrix renormalization group \cite{DMRG-White,Dukelsky-Pittel-RPP,Scholl-RMP}
(DMRG) scheme that represents a  very powerful and  general decimation prescription.
Currently, the DMRG algorithm is 
understood as an energy minimization within a class of low entanglement wavefunctions known as 
matrix product states  \cite{Scholl-AP,GChan}  (MPS) establishing an exciting link with quantum information
perpectives. \cite{Cirac} A very flexible entanglement encoding is also provided by the rapidly expanding research area 
of tensor network states \cite{TNPS-1,TNPS-2,TNPS-3} (TNS).

Variational principles also offer very powerful methods
to study Hubbard-like models. For example, the 
dynamical variational principle, \cite{DVP-1,DVP-2} expressed in the language of Green's
functions and self-energies, \cite{Fetter-W} provides us with the  
variational cluster approximation \cite{VCA-1} (VCA),  the 
dynamical impurity 
approximation \cite{DIA-1} (DIA) and the dynamical mean field theory \cite{DMFT-1} (DMFT).  
Within this context, the  
self-energy-functional theory \cite{SFT-1}
(SFT) has emerged as a  conceptual framework in which the 
VCA, DIA and DMFT, as well as  several extensions of them, can be 
specified by the choice of a reference system. In particular, the cluster extensions 
to DMFT have provided important
insights into the physics of the 2D Hubbard model in aspects such as
the Mott-Hubbard transition, the pseudogap in doped systems, and the
phase diagram itself.\cite{maier2005,stanescu2006} DMFT and its
cluster extensions are particularly valuable as they have been shown
to be complementary to  finite size 
simulations, \cite{maier2005,moukouri2001, huscroft2001,aryanpour2003} including ours.
Here, we also refer the interested reader to recent work \cite{Zgid}
 where a hierarchy of 
truncated configuration interaction (CI) expansions has been considered as a solver
for quantum impurity models and DMFT.

In the present  work, we explore an alternative avenue 
not only to describe  ground state properties of the 
2D Hubbard model
but also to access 
excitation
spectra which represent a basic
fingerprint of quantum mechanical correlations in the considered 
lattices. A first step in this direction, based
on symmetry-projected configuration mixing ideas originally employed  in 
microscopic 
nuclear structure 
theory, \cite{Carlo-review} was undertaken for the 1D 
Hubbard model \cite{Carlos-Hubbard-1D} and is extended in the 
present work to the 2D Hubbard Hamiltonian 
with periodic boundary conditions (PBC).

For a given
 single-electron space, we construct 
the most general unitary Hartree-Fock (HF)  transformation. \cite{rs,Blaizot-Ripka} Since 
this HF-transformation mixes all the spin and linear momentum quantum numbers of the 
single-electron basis states, the corresponding Slater determinant deliberately
breaks the original spin and translational symmetries of the  2D Hubbard Hamiltonian. Therefore, as such, our
symmetry-broken 
Slater determinant can be considered as a convenient mean-field starting point
enlarging the  space of trial wave functions. \cite{rs,Blaizot-Ripka}
We restore the broken  translational and spin symmetries with the help of linear  and 
angular momentum projection operators. This symmetry restoration  recovers the 
multi-determinantal character in our trial state keeping  good 
spin and linear momentum quantum numbers. The 
 Ritz variational principle 
\cite{rs,Blaizot-Ripka} is then applied to the projected energy, i.e., ours
is a variation-after-projection (VAP) scheme. This procedure 
provides us with the optimal (variational) representation 
of a ground state, with well defined spin and linear momentum quantum numbers, via
 a single  symmetry-projected configuration. Our VAP scheme is also very close 
 in spirit to  Projected Quasiparticle Theory \cite{PQT-reference-1,PQT-reference-2} (PQT) 
 and is 
 related to other variational approaches.\cite{Tomita-1,Tomita-2}   

%%%%%%%%%%%%%%%%%%%%%%%%%%%%%%%%%%%%%%%%%%%%%%%%%%%%%%%%%%%%%%%%%%%%%%%%%%%%%%%%%%%%%%%%%%%%%%%%%
%
%   FIGURE 1 OF THE PAPER
%
%%%%%%%%%%%%%%%%%%%%%%%%%%%%%%%%%%%%%%%%%%%%%%%%%%%%%%%%%%%%%%%%%%%%%%%%%%%%%%%%%%%%%%%%%%%%%%%%%
\begin{figure*}
\includegraphics[width=0.80\textwidth]{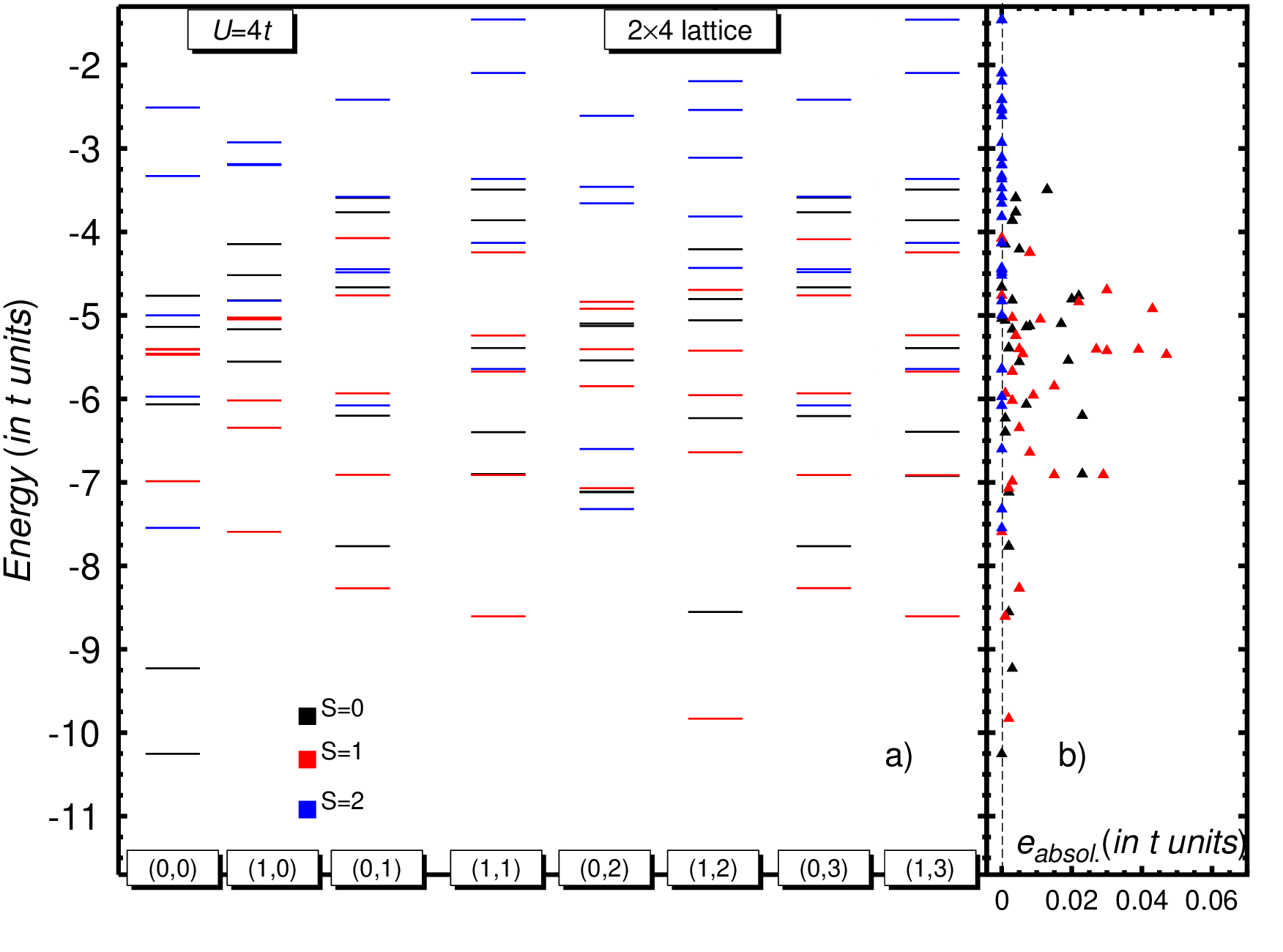} 
\caption{(Color online) The energy spectrum,  obtained via Eq.(\ref{diagonalizationfinal}), for
 the half-filled $2 \times 4$ lattice 
 at U=4t
 is shown in panel a). This spectrum can be hardly distinguished from the one 
obtained using an exact diagonalization (ED). Therefore, in panel b) the 
absolute errors  are plotted for each of the predicted 120 solutions.
For more details, see the main text. 
}
\label{2by4_lattice_spectrum_Ne8} 
\end{figure*}
%%%%%%%%%%%%%%%%%%%%%%%%%%%%%%%%%%%%%%%%%%%%%%%%%%%%%%%%%%%%%%%%%%%%%%%%%%%%%%%%%%%%%%%%%%%%%%%%%
%
%   END OF FIGURE 1 OF THE PAPER
%
%%%%%%%%%%%%%%%%%%%%%%%%%%%%%%%%%%%%%%%%%%%%%%%%%%%%%%%%%%%%%%%%%%%%%%%%%%%%%%%%%%%%%%%%%%%%%%%%%

In order to describe excited states with well defined quantum numbers, we construct a truncated basis consisting of a few 
(orthonormalized) symmetry-projected states throughout a chain of VAP calculations. This can be 
easily done, still with low computational cost, due to the simple structure of our  projected wave functions.
Finally, a further diagonalization of the 2D
Hubbard Hamiltonian is performed within such a basis. With this configuration mixing procedure 
we may account, in a similar fashion, for additional correlations in both
ground and excited states. In addition, our theoretical framework  can be used to
study important dynamical properties of the 2D Hubbard Hamiltonian like 
spectral functions. \cite{Dagotto-review,Szczepanski,Fetter-W}

In this paper we have three main goals. First, we present the methodology of a VAP
configuration mixing scheme, originally devised for the nuclear many-body problem, but 
not yet explored for the 2D Hubbard model. Therefore, in Sec. \ref{theory} we introduce 
our theoretical 
formalism. Symmetry restoration is described in 
Sec. \ref{formalism-H2D} while our configuration mixing scheme is outlined 
in Sec. \ref{formalism-H2D-excited}. For 
the reader's convenience, the key ingredients of our approximations are stressed 
in these two sections while, to make our presentation self-contained, more technical details 
can be found in appendices \ref{App-1}
and \ref{App-2}, respectively. Our second goal is to show how our theoretical framework can be used to 
access the spectral weight 
of states with different linear momentum quantum
numbers. To this end, the computation of hole and particle spectral 
functions
is briefly described in Sec. \ref{spectral-H2D} and more details are given in 
appendix \ref{App-3}. Our third goal is to  test the performance of our approximation for a selected set 
of illustrative examples. The 
results of our calculations 
 for the  half-filled $2 \times 4$, $4 \times 4$ and  $6 \times 6$  lattices  are 
 discussed in Sec. \ref{results}. There, we pay attention to the properties 
 of ground and excited states but also
 discuss  hole and particle spectral functions as well as the corresponding 
 density of states (DOS). In addition, in the case of the $4 \times 4$ lattice, we  consider
 doped systems with 14 and 15 electrons. Finally, Sec. \ref{Conclusions-work}  is 
devoted to the concluding 
remarks and work perspectives.

\section{Theoretical Framework}
\label{theory}
In what follows, we describe the theoretical 
framework used in the present study. First, symmetry
restoration and configuration mixing are presented in Secs. 
\ref{formalism-H2D} and \ref{formalism-H2D-excited}. The computation 
of spectral functions is briefly described in Sec. \ref{spectral-H2D}.

\subsection{Symmetry restoration  for the 2D Hubbard model}
\label{formalism-H2D}

We consider the following one-band version of the 2D
Hubbard Hamiltonian \cite{Hubbard-model_def1}

\begin{eqnarray} \label{HAM-HUB-2D}
\hat{H}_{Hub} &=& -t \sum_{{\bf{j}} {\sigma}} 
\Big(\hat{c}_{{\bf{j}}+{\bf{x}} \sigma}^{\dagger} \hat{c}_{{\bf{j}} \sigma} 
+ \hat{c}_{{\bf{j}}+{\bf{y}} \sigma}^{\dagger} \hat{c}_{{\bf{j}} \sigma}
+h.c.
\Big)
\nonumber\\
 &+& U
\sum_{{\bf{j}}}
\hat{c}_{{\bf{j}} \uparrow}^{\dagger}
\hat{c}_{{\bf{j}} \downarrow}^{\dagger}
\hat{c}_{{\bf{j}} \downarrow}
\hat{c}_{{\bf{j}} \uparrow}
\end{eqnarray}
where the first term represents the nearest-neighbor 
hopping  (t $>$ 0), with unit hopping vectors 
${\bf{x}}=(1,0)$ and  ${\bf{y}}=(0,1)$, and the second is  the repulsive on-site  interaction (U $>$ 0).
The operators $\hat{c}_{{\bf{j}} \sigma}^{\dagger}$ and $\hat{c}_{{\bf{j}} \sigma}$
create and destroy a particle with spin-projection 
$\sigma= \pm 1/2$ (also denoted as $\sigma= \uparrow, \downarrow$) along an arbitrary chosen 
quantization axis on a lattice site ${\bf{j}}$=($j_{x},j_{y}$). They satisfy 
the usual anticommutation relations for fermion operators.\cite{Blaizot-Ripka} Here, and in what follows, the 
lattice indices run as $j_{x}=1, \dots, N_{x}$ and $j_{y}=1, \dots, N_{y}$ with $N_{x}$ and $N_{y}$ being the number
of sites along the x and y directions, respectively. The total number of sites is given by $N_{sites}=N_{x} \times N_{y}$. 
We assume PBC, i.e., the sites $N_{i}+1$  and 1, with i=x,y, are
 identical. Furthermore, we
assume a lattice  spacing $\Delta$=1. 

Next, we apply the 2D Fourier transform

\begin{eqnarray} \label{Fourier-2D}
\hat{c}_{{\boldsymbol{\alpha}} \sigma}^{\dagger} =  
\frac{1}{\sqrt{N_{sites}}}
\sum_{{\bf{j}}}
e^{-i {\bf{k}}_{\alpha} {\bf{j}}}
\hat{c}_{{\bf{j}} \sigma}^{\dagger}
\end{eqnarray}
to obtain  operators   
with momentum 
${\bf{k}}_{\alpha}= \left(k_{{\alpha}_{x}},k_{{\alpha}_{y}}\right)
=
\left(\frac{2\pi {\alpha}_{x}}{N_{x}},\frac{2\pi {\alpha}_{y}}{N_{y}} \right)$. 
The Hamiltonian 
(\ref{HAM-HUB-2D}) can be easily written in terms of these new operators.
The quantum numbers 
${\alpha}_{i}$, with i=x,y, take the allowed values

\begin{eqnarray}  \label{values-momenta}
{\alpha}_{i}= -\frac{N_{i}}{2}+1, \dots ,\frac{N_{i}}{2} 
\end{eqnarray}
inside the Brillouin zone (BZ). \cite{Ashcroft-Mermin-book} Equivalently, they can take 
all integer values  between 
0 and $N_{i}-1$.

%%%%%%%%%%%%%%%%%%%%%%%%%%%%%%%%%%%%%%%%%%%%%%%%%%%%%%%%%%%%%%%%%%%%%%%%%%%%%%%%%%%%%%%%%%%%%%%%%%%%%%%%%%%%%
%
%    FIGURE 2 OF THE PAPER
%
%%%%%%%%%%%%%%%%%%%%%%%%%%%%%%%%%%%%%%%%%%%%%%%%%%%%%%%%%%%%%%%%%%%%%%%%%%%%%%%%%%%%%%%%%%%%%%%%%%%%%%%%%%%%%
\begin{figure*} 
\includegraphics[width=1.00\textwidth]{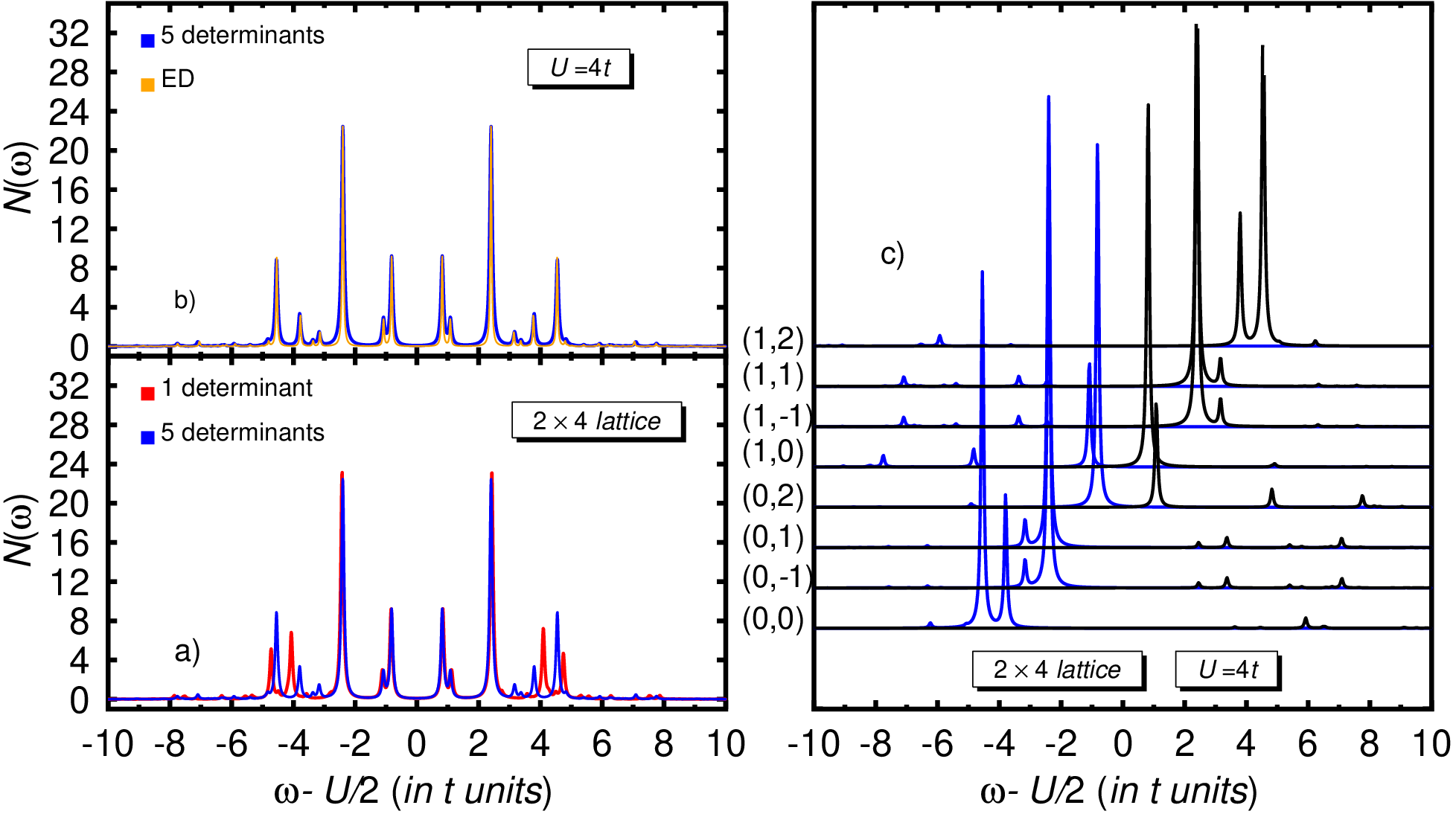} 
\caption{(Color online) The DOS ${\cal{N}}(\omega)$ [Eq.(\ref{densityofstates})]
for the half-filled $2 \times 4$ lattice at U=4t
is plotted in panel a) as a function of the 
shifted 
excitation energy $\omega-U/2$ (in t units). Results 
have been obtained by approximating the ($N_{e} \pm 1$)-electron systems  
[Eqs. (\ref{hole-wf-spectral}) and (\ref{particle-wf-spectral})]
with
 $n_{T}=1$ (red) and $n_{T}=5$ (blue) Slater determinants 
 out of Sec. \ref{formalism-H2D}. As can be observed from panel b) the 
 DOS obtained with  exact diagonalization (ED)
 and the one obtained using $n_{T}=5$ HF-transformations can hardly be 
 distinguished. The  
hole (blue) and particle (black) spectral functions, computed with $n_{T}=5$ HF-transformations, are 
plotted in panel c). A Lorentzian folding of width $\Gamma$=0.05t has been used.
}
\label{densitystates_halfilling_2by4_U4} 
\end{figure*}
%%%%%%%%%%%%%%%%%%%%%%%%%%%%%%%%%%%%%%%%%%%%%%%%%%%%%%%%%%%%%%%%%%%%%%%%%%%%%%%%%%%%%%%%%%%%%%%%%%%%%%%%%%%%%
%
%    END OF FIGURE 2 OF THE PAPER
%
%%%%%%%%%%%%%%%%%%%%%%%%%%%%%%%%%%%%%%%%%%%%%%%%%%%%%%%%%%%%%%%%%%%%%%%%%%%%%%%%%%%%%%%%%%%%%%%%%%%%%%%%%%%%%

In the HF-approximation, the 
ground state of an $N_{e}$-electron system is  represented 
by a Slater determinant $|{\cal{D}} \rangle = \prod_{i=1}^{N_{e}} \hat{b}_{h}^{+} | 0 \rangle$ in which the energetically 
lowest $N_{e}$ single-electron states
(hole states $h$, $h^{'}$, \dots )
 are occupied while the 
remaining $2N_{sites}-N_{e}$ states 
(particle states $p$, $p^{'}$, \dots)
are empty. The HF-quasiparticle operators  are given by

\begin{eqnarray} \label{HF-transformation}
\hat{b}_{a}^{\dagger} = \sum_{{\boldsymbol{\alpha}} {\sigma}}
{\cal{D}}_{{\boldsymbol{\alpha}} \sigma,a}^{*}
\hat{c}_{{\boldsymbol{\alpha}} \sigma}^{\dagger}
\end{eqnarray}
where ${\cal{D}}$ is a general  $2N_{sites} \times 2N_{sites}$ unitary transformation. \cite{rs,Blaizot-Ripka} In Eq. (\ref{HF-transformation}) $a$ is a shorthand 
notation for the set $(a_{x},a_{y}, {\sigma}_{a})$. The transformation 
(\ref{HF-transformation}) mixes all the linear momentum 
quantum numbers 
as well as the spin projection  of the  states 
(\ref{Fourier-2D}).  As a consequence, $|{\cal{D}} \rangle$  deliberately breaks  rotational (in spin space) and 
translational invariances. To restore the spin quantum numbers
we explicitly use the projection operator

\begin{eqnarray} \label{PROJ-S}
\hat{P}_{\Sigma {\Sigma}^{'}}^{S} =
\frac{2S+1}{8 {\pi}^{2}} \int d \Omega {\cal{D}}_{\Sigma {\Sigma}^{'}}^{S *} (\Omega) R_{S}(\Omega)
\end{eqnarray}
where $ R_{S}(\Omega)= e^{-i \alpha \hat{S}_{z}} e^{-i \beta \hat{S}_{y}} e^{-i \gamma \hat{S}_{z}}$ is
 the rotation operator in spin space, $\Omega = \left(\alpha, \beta, \gamma \right)$ stands for the set of Euler 
angles and ${\cal{D}}_{\Sigma {\Sigma}^{'}}^{S}(\Omega)$ are Wigner functions. \cite{Edmonds} The form  
(\ref{PROJ-S}) has been frequently used for total angular momentum projection in nuclear physics. 
\cite{rs,Carlo-review} This form has also  been adopted in the study of the 1D Hubbard model \cite{Carlos-Hubbard-1D} and 
more recently within 
PQT in quantum chemistry. \cite{PQT-reference-1,PQT-reference-2} 

The  linear momenta, $k_{{\xi}_{x}}$ 
and 
$k_{{\xi}_{y}}$, are  restored 
with the projector 

\begin{eqnarray} \label{projector-LM}
\hat{C}({\boldsymbol{\xi}})
=
\frac{1}{N_{sites}} 
\sum_{{\bf{j}}}
e^{i \left(j_{x}+j_{y} \right) \hat{P} }
e^{-i {\bf{k}}_{\xi} {\bf{j}}}
\end{eqnarray}
where $\hat{P} = \displaystyle \sum_{{\boldsymbol{\alpha}} \sigma}
\left( k_{{\alpha}_{x}} + k_{{\alpha}_{y}} \right)
\hat{c}_{{\boldsymbol{\alpha}} \sigma}^{\dagger}
\hat{c}_{{\boldsymbol{\alpha}} \sigma}$
is the generator of the considered  
lattice translations. Note that this operator 
neither has  vector character nor corresponds to the 
true linear momentum operator. It is associated with the quasi-momentum resulting from
translational invariance of the lattice. We will refer to it, however, as linear momentum
for simplicity. The projector 
(\ref{projector-LM}) represents the 2D limit of the 
general operator restoring 
Galilei invariance.  \cite{Carlo-review,Rayner-Carlo-CM-1,Rayner-Carlo-CM-2} Note that, at variance with 
atomic nuclei, lattice systems can have solutions with linear momenta different from zero.

%%%%%%%%%%%%%%%%%%%%%%%%%%%%%%%%%%%%%%%%%%%%%%%%%%%%%%%%%%%%%%%%%%%%%%%%%%%%%%%%%%%%%%%%%%%%%%%%%%%%%%%%%%%%
%
% FIGURE 3 OF THE PAPER 
%
%%%%%%%%%%%%%%%%%%%%%%%%%%%%%%%%%%%%%%%%%%%%%%%%%%%%%%%%%%%%%%%%%%%%%%%%%%%%%%%%%%%%%%%%%%%%%%%%%%%%%%%%%%%%
\begin{figure*} 
\includegraphics[width=0.70\textwidth]{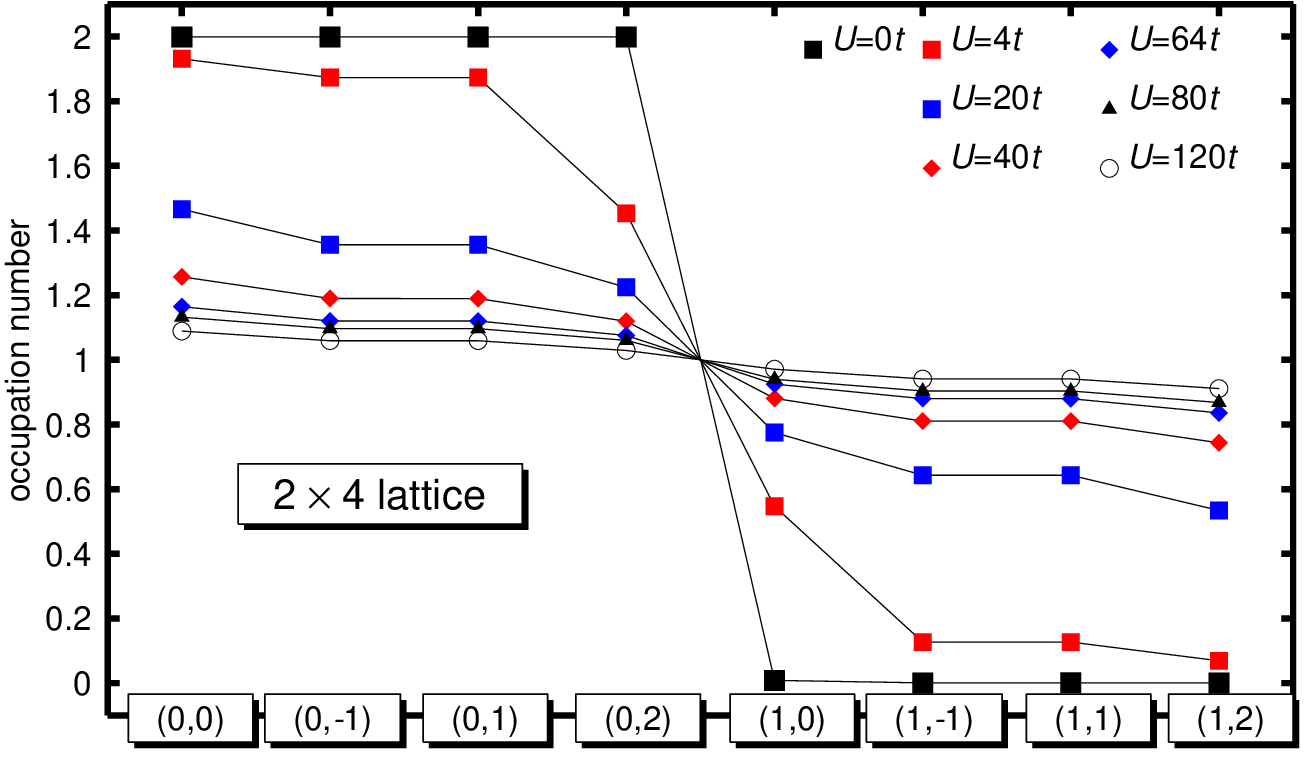} 
\caption{(Color online) Occupation numbers [Eq.(\ref{occupation-numbers})]  of the basis
states in the ground state of the half-filled $2 \times 4$ lattice 
are plotted for various $U$ strengths. 
}
\label{occupations_halfilling_2by4_U4} 
\end{figure*}
%%%%%%%%%%%%%%%%%%%%%%%%%%%%%%%%%%%%%%%%%%%%%%%%%%%%%%%%%%%%%%%%%%%%%%%%%%%%%%%%%%%%%%%%%%%%%%%%%%%%%%%%%%%%
%
% END OF FIGURE 3 OF THE PAPER 
%
%%%%%%%%%%%%%%%%%%%%%%%%%%%%%%%%%%%%%%%%%%%%%%%%%%%%%%%%%%%%%%%%%%%%%%%%%%%%%%%%%%%%%%%%%%%%%%%%%%%%%%%%%%%%

In what 
follows, we
introduce the  shorthand notation
$\Theta = (S,{\boldsymbol{\xi}})$ for the 
set of (symmetry) quantum 
numbers $(S,{\xi}_{x},{\xi}_{y})$, i.e.,  $\hat{P}_{\Sigma {\Sigma}^{'}}^{S }\hat{C}({\boldsymbol{\xi}}) = \hat{P}_{\Sigma {\Sigma}^{'}}^{\Theta}$. We
 then use the following
symmetry-projected  wave function

\begin{eqnarray} \label{SP-WF-VAMPIR}
| {\cal{D}}; \Theta; \Sigma \rangle
=
\sum_{{\Sigma}^{'}=-S}^{S} f_{{\Sigma}^{'}}^{ \Theta}
\hat{P}_{\Sigma {\Sigma}^{'}}^{\Theta} 
| {\cal{D}} \rangle
\end{eqnarray} 
where $f_{{\Sigma}^{'}}^{ \Theta}$ are variational parameters. Note that, through the action of the projection operator $\hat{P}_{\Sigma {\Sigma}^{'}}^{\Theta}$, the 
multi-determinantal character of the state characterized by the quantum 
numbers $\Theta$ and ${\Sigma}^{'}$ is recovered and written in terms of the quantum numbers $\Theta$ and $\Sigma$. \cite{rs}
In practice, the integration over the set of Euler angles  
in Eq.(\ref{PROJ-S}) is 
discretized. For the integrals in $\alpha$ and $\gamma$ we have used 8 grid points whereas 
for the $\beta$-integration we have used 16 points. Therefore, the total number of grid points to be used 
for the projection operator $\hat{P}_{\Sigma {\Sigma}^{'}}^{\Theta}$ is $1024 \times N_{sites}$.

For a given symmetry $\Theta$, the energy (independent of $\Sigma$) 
associated with the state (\ref{SP-WF-VAMPIR}) 

\begin{eqnarray} \label{energy-vampir}
E^{\Theta} = 
\frac{
f^{ \Theta \dagger}
{\cal{H}}^{ \Theta}
f^{ \Theta}
}
{
f^{ \Theta \dagger}
{\cal{N}}^{ \Theta}
f^{ \Theta}}
\end{eqnarray} 
is 
given 
in terms of the  $(2S+1)\times (2S+1)$ Hamiltonian 
${\cal{H}}_{{\Sigma} {\Sigma}^{'}}^{ \Theta}=
\langle {\cal{D}} | \hat{H}_{Hub} \hat{P}_{\Sigma {\Sigma}^{'}}^{\Theta}  | {\cal{D}} \rangle
$ and 
norm 
${\cal{N}}_{{\Sigma} {\Sigma}^{'}}^{ \Theta}
= \langle {\cal{D}} | \hat{P}_{\Sigma {\Sigma}^{'}}^{\Theta}  | {\cal{D}} \rangle
$ matrices (see appendix \ref{App-1}). It has to be 
minimized with respect to  the coefficients  
$f^{ \Theta}$ and the HF-transformation
${\cal{D}}$. The variation with respect to the mixing coefficients  yields the 
following 
generalized eigenvalue equation

\begin{eqnarray} \label{HW-1}
\left({\cal{H}}^{\Theta}
- E^{ \Theta}
{\cal{N}}^{\Theta }
\right)
f^{\Theta} = 0
\end{eqnarray} 
with the constraint $f^{\Theta \dagger} {\cal{N}}^{\Theta} f^{\Theta} = 1_{2S+1}$
ensuring the orthogonality of the solutions. The 
unrestricted minimization of the 
energy (\ref{energy-vampir}) with respect to the underlying 
HF-transformation ${\cal{D}}$ can be carried out  via the Thouless
theorem. \cite{rs,Carlo-review}  The corresponding 
variational equations assume the form 

\begin{eqnarray} \label{gradient}
M_{{\Theta}}^{ -1 \dagger} G^{\Theta}  L_{\Theta} =0
\end{eqnarray}
with

\begin{eqnarray} \label{local-gradient}
G_{ph}^{\Theta}=
\Big[ 
f^{ \Theta \dagger}
\left(
{\cal{K}}^{ \Theta}
-E^{ \Theta} {\cal{R}}^{ \Theta}
\right)
f^{ \Theta}
\Big]_{ph}
\end{eqnarray}
Here, the $N_{e} \times N_{e}$ and $(2N_{sites}-N_{e}) \times (2N_{sites}-N_{e})$
matrices $L_{{\Theta}}$ and $M_{{\Theta}}$ 
are obtained via the Cholesky decompositions. \cite{Carlo-review,Rayner-Carlo-CM-1}
The particle-hole kernels  
$
{\cal{K}}_{\Sigma {\Sigma}^{'}}^{ \Theta;ph}
= \langle {\cal{D}}| \hat{H}_{Hub} \hat{P}_{\Sigma {\Sigma}^{'}}^{\Theta}
{\hat{b}}^{\dagger}_{p} {\hat{b}}_{h}
| {\cal{D}} \rangle
$ and  
${\cal{R}}_{\Sigma {\Sigma}^{'}}^{\Theta;ph}
=
\langle {\cal{D}} |  \hat{P}_{\Sigma {\Sigma}^{'}}^{\Theta}
{\hat{b}}^{\dagger}_{p} {\hat{b}}_{h}
| {\cal{D}} \rangle$ are given in appendix \ref{App-2}.
It should be stressed that, for a given symmetry $\Theta$, we only retain
 the energetically lowest solution of Eqs.(\ref{HW-1}) 
and (\ref{gradient}). Both the HF-transformation 
${\cal{D}}$ and the mixing coefficients $f^{\Theta}$ are essentially complex, therefore
one needs to minimize $n_{var}=2(2N_{sites}-N_{e})\times N_{e}+ 4S$
real variables. We use a quasi-Newton method for such a minimization. \cite{quasi-Newton-1,quasi-Newton-2} The 
variational procedure already described is known in nuclear structure physics as the VAMPIR (i.e., Variation
After Mean field Projection In Realistic model spaces). \cite{Carlo-review} Note, that particle 
number projection \cite{rs} is not carried out in the present study since the considered Slater determinants
conserve the number of electrons.

%%%%%%%%%%%%%%%%%%%%%%%%%%%%%%%%%%%%%%%%%%%%%%%%%%%%%%%%%%%%%%%%%%%%%%%%%%%%%%%%%%%%%%%%%%%%%%%%%%%%%%%%%%%%%%
%
%  FIGURE 4 OF THE PAPER
%
%%%%%%%%%%%%%%%%%%%%%%%%%%%%%%%%%%%%%%%%%%%%%%%%%%%%%%%%%%%%%%%%%%%%%%%%%%%%%%%%%%%%%%%%%%%%%%%%%%%%%%%%%%%%%%
\begin{figure*} 
\includegraphics[width=1.0\textwidth]{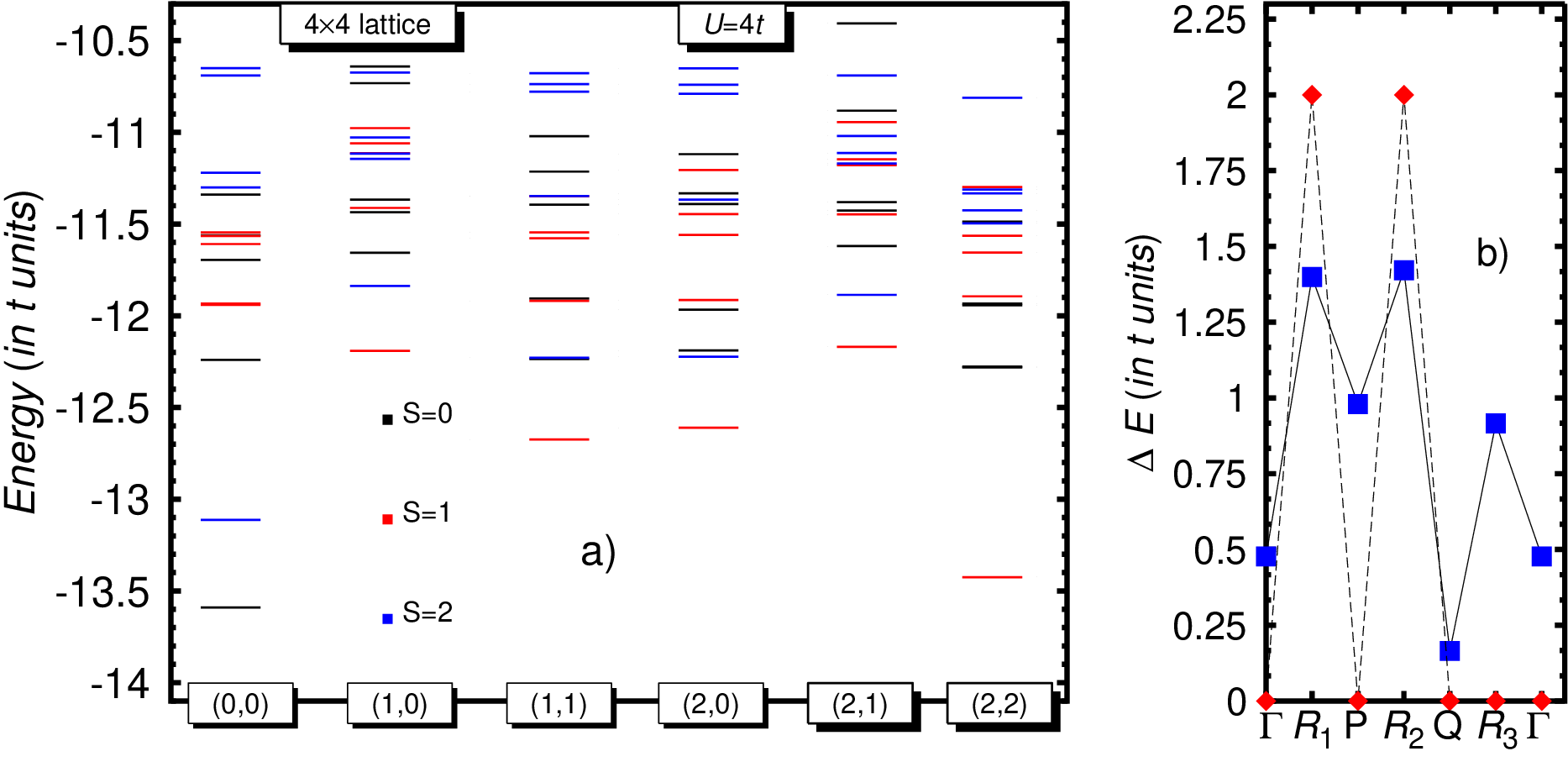} 
\caption{(Color online) The energy spectrum,  obtained via Eq.(\ref{diagonalizationfinal}), for
 the half-filled $4 \times 4$ lattice at U=4t is shown in panel a). In panel b), the excitation energies 
 from the ground state to the lowest-lying
 S=1 and S=2 states 
from panel a) are plotted as functions 
 of the linear momentum quantum numbers $\Gamma=(0,0)$, $R_{1}=(1,0)$, $P=(2,0)$, $R_{2}=(2,1)$, $Q=(2,2)$, and 
$R_{3}=(1,1)$, respectively. In addition to U=4t (blue boxes), results for U=0t (red diamonds) 
are also included for comparison.  
}
\label{4by4_lattice_spectrum_Ne16}
\end{figure*}
%%%%%%%%%%%%%%%%%%%%%%%%%%%%%%%%%%%%%%%%%%%%%%%%%%%%%%%%%%%%%%%%%%%%%%%%%%%%%%%%%%%%%%%%%%%%%%%%%%%%%%%%%%%%%%
%
%  END OF FIGURE 4 OF THE PAPER
%
%%%%%%%%%%%%%%%%%%%%%%%%%%%%%%%%%%%%%%%%%%%%%%%%%%%%%%%%%%%%%%%%%%%%%%%%%%%%%%%%%%%%%%%%%%%%%%%%%%%%%%%%%%%%%%

\subsection{Symmetry-projected configuration mixing  for the 2D Hubbard model}
\label{formalism-H2D-excited}

An accurate description of  excited states in a many-fermion system is much more 
difficult even when one is usually interested  in just a small fraction 
of the low-lying spectrum. Here, the main difficulty in the optimization of excited states 
is ensuring orthogonality among them and with respect to the ground state. For this, we simply use 
a Gram-Schmidt orthogonalization. Our goal in this section is to construct, throughout
a chain of VAP calculations, a basis of a few (orthonormalized)  states 
with well defined  quantum
numbers $\Theta$.

Suppose we have generated 
a ground state solution $ | {\phi}^{1} \rangle = | {\cal{D}}; \Theta; \Sigma \rangle$ 
out of Eqs. (\ref{HW-1}) and 
(\ref{gradient}) in Sec. \ref{formalism-H2D}. Then we write the first 
excited state wave function as

\begin{eqnarray} \label{example-excvampir}
| {\varphi}_{2} \rangle = {\beta}_{1}^{2} |{\phi}^{1} \rangle + {\beta}_{2}^{2} | {\phi}^{2} \rangle
\end{eqnarray}
where $| {\phi}^{2} \rangle$ has a form similar to Eq.(\ref{SP-WF-VAMPIR}) but 
with different  coefficients $f^{2 \Theta}$ and 
underlying HF-transformation ${\cal{D}}^{2}$. The label 2 distinguishes them from 
the ones (i.e., $f^{1 \Theta}$ and ${\cal{D}}^{1}$) corresponding to the reference 
ground state we already have. Both
${\beta}_{1}^{2}$ and ${\beta}_{2}^{2}$  can be obtained by requiring that
$\langle {\phi}^{1} |  {\varphi}_{2} \rangle=0$ and 
$\langle {\varphi}_{2} |  {\varphi}_{2} \rangle=1$. They are given in
terms of the projector (i.e., $\hat{S}_{1}= \hat{S}_{1}^{2}$) 

\begin{eqnarray}
\hat{S}_{1}= \frac{ |{\phi}^{1} \rangle \langle {\phi}^{1} |}{\langle {\phi}^{1} | {\phi}^{1} \rangle }   
\end{eqnarray}
as follows 

\begin{eqnarray}
{\beta}_{2}^{2} &=& \langle {\phi}^{2} 
| \left(1 - \hat{S}_{1}  \right)
| {\phi}^{2} \rangle^{-1/2}
\nonumber\\
{\beta}_{1}^{2} &=& - \frac{ \langle {\phi}^{1} | {\phi}^{2} \rangle}{\langle {\phi}^{1} | {\phi}^{1} \rangle }  
{\beta}_{2}^{2}
\end{eqnarray}

The first excited state is obtained
varying the energy functional for (\ref{example-excvampir}) with respect to 
$f^{2 \Theta}$ and 
${\cal{D}}^{2}$. For the second excited state, we introduce a new state
$| {\phi}^{3} \rangle$, again with the same  form as in Eq.(\ref{SP-WF-VAMPIR}), and write

\begin{eqnarray} \label{example-excvampir-2}
| {\varphi}_{3} \rangle = {\beta}_{1}^{3} |{\phi}^{1} \rangle 
+ {\beta}_{2}^{3} | {\phi}^{2} \rangle
+ {\beta}_{3}^{3} | {\phi}^{3} \rangle
\end{eqnarray}
with coefficients ${\beta}_{1}^{3}$, ${\beta}_{2}^{3}$ and 
${\beta}_{3}^{3}$ such that $| {\varphi}_{3} \rangle$ is orthogonal
to the previous solutions $| {\varphi}_{1} \rangle= | {\phi}^{1} \rangle$
[Eq.(\ref{SP-WF-VAMPIR})]
and $| {\varphi}_{2} \rangle$ [Eq.(\ref{example-excvampir})] as well as 
$\langle {\varphi}_{3} |  {\varphi}_{3} \rangle=1$. The second 
excited state is obtained
varying the energy functional for (\ref{example-excvampir-2}) with respect to 
$f^{3 \Theta}$ and 
${\cal{D}}^{3}$. Let us have a more 
general situation in which, by successive variation, $i=1, \dots ,m-1$ orthonormalized solutions (for example, $| {\varphi}_{1} \rangle$ and 
$| {\varphi}_{2} \rangle$)  

\begin{eqnarray} \label{cuco-1}
|{\varphi}_{i} \rangle= \sum_{j=1}^{i} |{\phi}^{j} \rangle {\beta}_{j}^{i} 
\end{eqnarray}
are already at our disposal. Each of the states  $| {\phi}^{j} \rangle$ in 
(\ref{cuco-1}) has the same form as (\ref{SP-WF-VAMPIR}). One then writes the ansatz for  
the $m$th  state wave function (for example, $| {\varphi}_{3} \rangle$) as

\begin{eqnarray} \label{cuco-2}
| {\varphi}_{m} \rangle= \sum_{j=1}^{m-1} |{\phi}^{j} \rangle {\beta}_{j}^{m} + |{\phi}^{m} \rangle {\beta}_{m}^{m}
\end{eqnarray}
with $|{\phi}^{m} \rangle$ having again the form (\ref{SP-WF-VAMPIR}). Requiring 
orthonormalization with respect to all the previous $m-1$ solutions (\ref{cuco-1}) the 
coefficients  ${\beta}_{m}^{m}$ and ${\beta}_{j}^{m}$ in Eq.(\ref{cuco-2}) read 

\begin{eqnarray} \label{cuco-3}
{\beta}_{m}^{m} &=& \langle {\phi}^{m} 
| \left( 1 - \hat{S}_{m-1} \right)
| {\phi}^{m} \rangle^{-1/2} 
\nonumber\\
{\beta}_{j}^{m} &=& - \sum_{k=1}^{m-1} 
\frac{
\langle {\phi}^{k} | {\phi}^{m} \rangle 
}
{
\langle {\phi}^{j} | {\phi}^{k} \rangle
} 
{\beta}_{m}^{m}
\end{eqnarray}
in terms of the projector (i.e., $\hat{S}_{m-1}= \hat{S}_{m-1}^{2}$)

\begin{eqnarray}
\hat{S}_{m-1}  
= \sum_{j,k=1}^{m-1}
\frac{
| {\phi}^{j} \rangle
\langle {\phi}^{k} |}
{
\langle {\phi}^{j} | {\phi}^{k} \rangle
}
\end{eqnarray}

%%%%%%%%%%%%%%%%%%%%%%%%%%%%%%%%%%%%%%%%%%%%%%%%%%%%%%%%%%%%%%%%%%%%%%%%%%%%%%%%%%%%%%%%%%%%%%%%%%%%%%%%%%%%%
%
% FIGURE 5 OF THE PAPER
%
%%%%%%%%%%%%%%%%%%%%%%%%%%%%%%%%%%%%%%%%%%%%%%%%%%%%%%%%%%%%%%%%%%%%%%%%%%%%%%%%%%%%%%%%%%%%%%%%%%%%%%%%%%%%%
\begin{figure*} 
\includegraphics[width=1.00\textwidth]{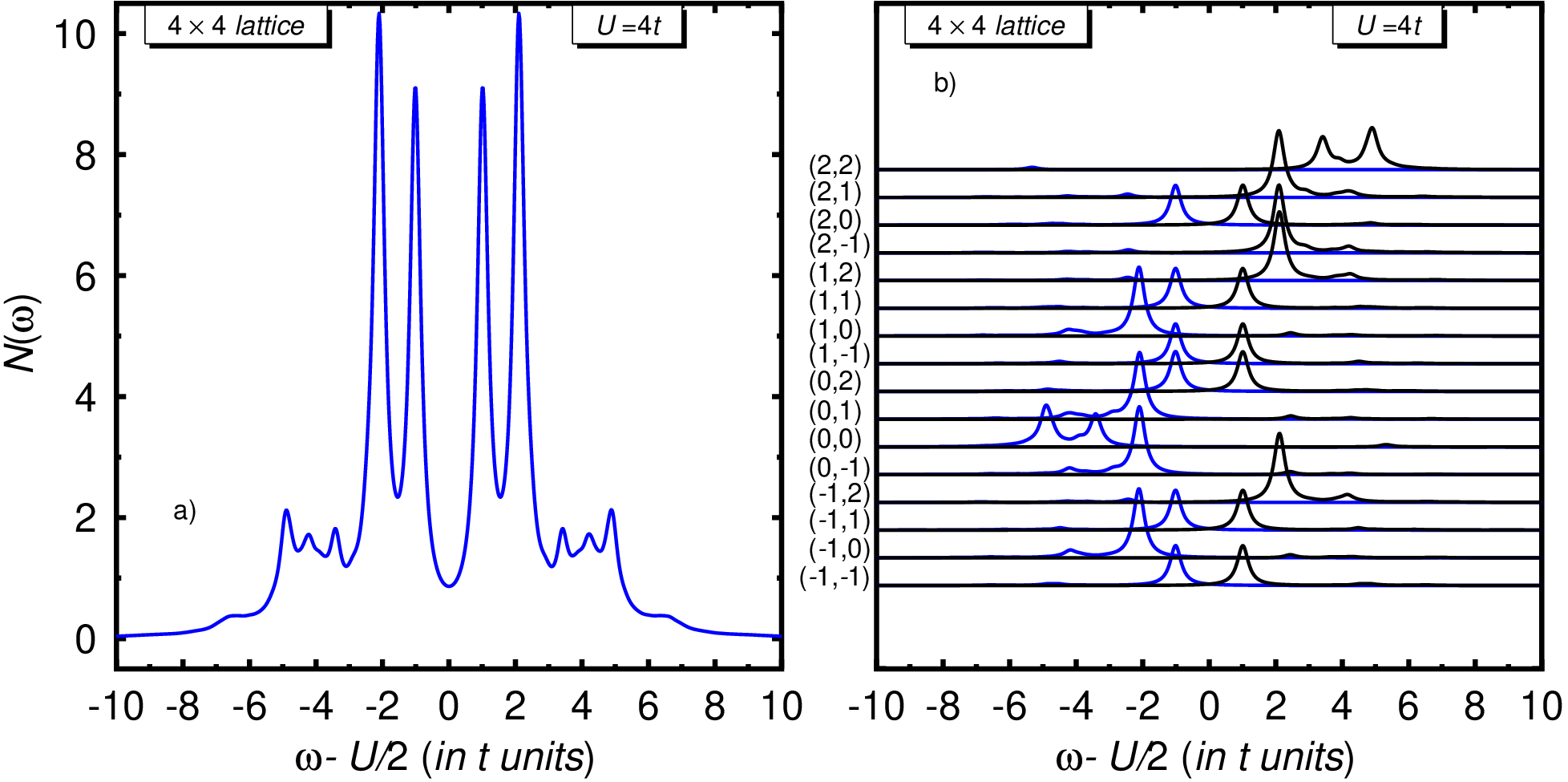} 
\caption{(Color online) The DOS ${\cal{N}}(\omega)$ [Eq.(\ref{densityofstates})]
for the half-filled $4 \times 4$ lattice at U=4t
is plotted in panel a) as a function of the shifted 
excitation energy $\omega-U/2$ (in t units). Results 
have been obtained by approximating the ($N_{e} \pm 1$)-electron systems  
[Eqs. (\ref{hole-wf-spectral}) and (\ref{particle-wf-spectral})]
with
 $n_{T}=5$ HF-determinants. Hole (blue) and particle (black) spectral functions, are 
displayed in panel b). A Lorentzian folding of width  $\Gamma$=0.2t has been used.
}
\label{densitystates_halfilling_4by4_U4} 
\end{figure*} 
%%%%%%%%%%%%%%%%%%%%%%%%%%%%%%%%%%%%%%%%%%%%%%%%%%%%%%%%%%%%%%%%%%%%%%%%%%%%%%%%%%%%%%%%%%%%%%%%%%%%%%%%%%%%%
%
% FIGURE 5 OF THE PAPER
%
%%%%%%%%%%%%%%%%%%%%%%%%%%%%%%%%%%%%%%%%%%%%%%%%%%%%%%%%%%%%%%%%%%%%%%%%%%%%%%%%%%%%%%%%%%%%%%%%%%%%%%%%%%%%%

The energy for the state (\ref{cuco-2}) takes the form

\begin{eqnarray} \label{ojo-ojo}
E^{m \Theta} =\frac{ 
f^{m \Theta \dagger}
{\cal{H}}^{m \Theta}
f^{m \Theta}
}
{
f^{m \Theta \dagger}
{\cal{N}}^{m \Theta}
f^{m \Theta}
}
\end{eqnarray}
with kernels ${\cal{H}}_{\Sigma {\Sigma}^{'}}^{m \Theta}$
and ${\cal{N}}_{\Sigma {\Sigma}^{'}}^{m \Theta}$ accounting for the fact
that $m$-1 linearly independent solutions 
have been removed from the variational space. Their expressions are 
slightly more involved \cite{Carlo-review} 
than the ones required in Eq.(\ref{energy-vampir})
but still
straightforward. They  require the knowledge of the 
symmetry-projected matrix elements 
between two different Slater determinants 
$ | {\cal{D}}^{i} \rangle$ and $ | {\cal{D}}^{k} \rangle$
(see appendix \ref{App-1}). The variation
of the energy (\ref{ojo-ojo}) with respect to 
$f^{m \Theta}$
 yields  an   equation similar
to (\ref{HW-1}) with the constraint 
   $f^{m \Theta \dagger} {\cal{N}}^{m \Theta} f^{m \Theta} = 1_{2S+1}$. The unrestricted minimization of the energy (\ref{ojo-ojo})
with respect to   ${\cal{D}}^{m}$, via the 
Thouless theorem,  leads to  variational
equations similar to (\ref{gradient}) but with
kernels  
${\cal{K}}_{\Sigma {\Sigma}^{'}}^{m \Theta; ph}$
and ${\cal{R}}_{\Sigma {\Sigma}^{'}}^{m \Theta; ph}$
that require
symmetry-projected particle-hole matrix elements 
between two different Slater determinants 
$ | {\cal{D}}^{i} \rangle$ and $ | {\cal{D}}^{k} \rangle$
(see appendix \ref{App-2}).

The procedure  outlined in this section is  known in  nuclear structure
physics
 as 
EXCITED VAMPIR. \cite{Carlo-review} It provides
 a  (truncated) basis of m (orthonormalized) states
$| {\varphi}_{j} \rangle$, with a well defined symmetry $\Theta$,
still keeping  low computational cost. This is doable due to the simple 
structure of the projected states defining such a basis in combination with a fast minimization 
scheme. \cite{quasi-Newton-1,quasi-Newton-2}  Our method can also be extended to use 
general Hartree-Fock-Bogoliubov (HFB) transformations. \cite{rs,Carlo-review,PQT-reference-1} However, this
 requires 
an additional projection of the
 particle number, which increases the  numerical effort
by about one order of magnitude and has hence not been
  used in the present paper.

It should be noticed that the ground state 
$| {\varphi}_{1} \rangle$ [Eq.(\ref{SP-WF-VAMPIR})] is written as a projection operator
acting on a single determinant, the first excited state $| {\varphi}_{2} \rangle$
[Eq.(\ref{example-excvampir})]
as a projection operator acting on two
determinants, and so on. Because this allows excited state wave functions to be described at a higher
level of quality than is the ground state wave function, our final step is to diagonalize the 
2D Hubbard Hamiltonian in the basis of the states $| {\varphi}_{j} \rangle$.

\begin{eqnarray} \label{diagonalizationfinal}
\sum_{j=1}^{m} \Big[ \langle {\varphi}_{i} | \hat{H}_{Hub} | {\varphi}_{j} \rangle 
- {\epsilon}_{\alpha}^{\Theta} 
{\delta}_{ij} \Big] C_{j \alpha}^{\Theta} = 0
\end{eqnarray}

For ground and excited states, the resulting wave functions 

\begin{eqnarray}
| {\Omega}_{\alpha}^{\Theta} \rangle= \sum_{\alpha} C_{j \alpha}^{\Theta} | {\varphi}_{j} \rangle
\end{eqnarray}
may account for more correlations than the description based on a single symmetry-projected configuration
discussed in Sec. \ref{formalism-H2D}. In the present work, as a first step, we have restricted 
ourselves to test the performance of our approximation with $m$=5  
(orthonormalized) states. As we will see, this turns out to be a reasonable starting point 
for, at least, a qualitative description of the considered lattices.

An interesting issue is the 
evolution of the energy of each state with the number $m$ of transformations 
included in the prescription described in this section. We observe
that, for the lattices considered in the present  study, the energy of the ground  and the first
couple of excited states remains unchanged  when $m$ goes from 1 to 5
(the changes in the energy per site are of the order $10^{-4}$).
This is partly because the main correlations have already been accounted 
for with a single symmetry-projected 
determinant. Therefore, the excited configurations obtained constitute reasonably good approximations 
to the true excited states of the considered system. We produce $m$=5 
symmetry-projected determinants in order to obtain the low-lying spectrum.
For the systems considered in this work, these 
states turn out to be weakly coupled through the 
Hamiltonian. However, this cannot be anticipated {\it{a priori}} and the  
diagonalization  Eq.(\ref{diagonalizationfinal}) should always be carried out.
Preliminary results for larger square lattices (i.e., $8 \times 8$ and 
$10 \times 10$) as well as for other Hamiltonians 
(i.e., the $t-t^{'}-U$ and $t-t^{'}-t^{"}-U$ Hubbard models) indicate that 
there are cases in 
which the diagonalization Eq.(\ref{diagonalizationfinal}) brings a sizeable amount of additional
correlations. 

\subsection{Hole and particle spectral functions}
\label{spectral-H2D}
Let us assume 
that for an even number $N_{e}$ of electrons
we already have the ground state wave function
$| {\cal{D}}^{1}; {\Theta}^{0}; \Sigma=0 \rangle$, out of the calculations 
described in Sec. \ref{formalism-H2D}. Since for all the considered lattices with 
an even number of electrons the ground state has spin S=0, but not 
neccessarily linear momenta zero, we write its 
quantum numbers as  ${\Theta}^{0}= (0,{\boldsymbol{\xi}}^{0})$.
Usually, spectral functions are computed within a Green's function 
perspective. \cite{Fetter-W} The key point is then to approximate
the ground states of the $(N_{e} \pm 1)$-electron systems by a 
suitable ansatz. In the present study,  we 
 approximate \cite{Carlos-Hubbard-1D} the ground state of the ($N_{e}$-1)-electron system, with 
the 
symmetry ${\Theta}^{-}= (S=1/2,{\boldsymbol{\xi}}^{-})$, by

\begin{eqnarray} \label{hole-wf-spectral}
| h_{1};{\Theta}^{-} ; \sigma  \rangle &=& 
\sum_{i h {\sigma}^{'}}
f_{i h {\sigma}^{'},h_{1}}^{{\Theta}^{-}}
\hat{P}^{{\Theta}^{-}}_{ \sigma {\sigma}^{'}}
\hat{b}_{h}({\cal{D}}^{i}) | {\cal{D}}^{i} \rangle
\end{eqnarray}
where the index i runs as $i=1, \dots, n_{T}$, the  hole index h as $h=1, \dots, N_{e}$ 
and ${\sigma}^{'}= \pm 1/2$. In Eq.(\ref{hole-wf-spectral}), we write 
$\hat{b}_{h}({\cal{D}}^{i})$ to explicitly indicate that holes are made on $n_{T}$ different 
Slater determinants. The determinants $|{\cal{D}}^{1} \rangle$ and $|{\cal{D}}^{i} \rangle$ 
correspond to the ground and lowest energy ($i=2, \dots, n_{T}$) states   
obtained for  the $N_{e}$-electron system out of the 
calculations described in Sec. \ref{formalism-H2D}. 
In the present study, we have restricted ourselves to a maximum 
of $n_{T}=5$ HF-transformations. The coefficients $f^{{\Theta}^{-}}$
in Eq.(\ref{hole-wf-spectral})
are obtained by solving the  equation

\begin{eqnarray} \label{HW-holes-H}
\left( {\cal{H}}^{{\Theta}^{-}} -
 E_{ h_{1}}^{{\Theta}^{-}}
 {\cal{N}}^{{\Theta}^{-}} \right) 
 f^{{\Theta}^{-}}= 0
\end{eqnarray}
that yields $2n_{T}N_{e}$ hole solutions $h_{1}$ with
 energies $E_{h_{1}}^{{\Theta}^{-}}$. With all the previous
  ingredients, one can 
compute \cite{Edmonds,Rayner-Carlo-CM-1,Rayner-Carlo-CM-2} the 
 hole spectral function as  
 $S_{h_{1}}({\boldsymbol{\xi}}^{-}, \delta \epsilon_{h_{1}})
 =
 |\langle h_{1}; {\Theta}^{-} || 
 {\hat{c}}_{{\boldsymbol{\xi}}^{0}-{\boldsymbol{\xi}}^{-}} || {\cal{D}}^{1}; {\Theta}^{0} \rangle |^{2}
 $
in terms of the reduced matrix element 

%%%%%%%%%%%%%%%%%%%%%%%%%%%%%%%%%%%%%%%%%%%%%%%%%%%%%%%%%%%%%%%%%%%%%%%%%%%%%%%%%%%%%%%%%%%%%%%%%%%%
%
%  FIGURE 6 OF THE PAPER
%
%%%%%%%%%%%%%%%%%%%%%%%%%%%%%%%%%%%%%%%%%%%%%%%%%%%%%%%%%%%%%%%%%%%%%%%%%%%%%%%%%%%%%%%%%%%%%%%%%%% 
\begin{figure*} 
\includegraphics[width=1.00\textwidth]{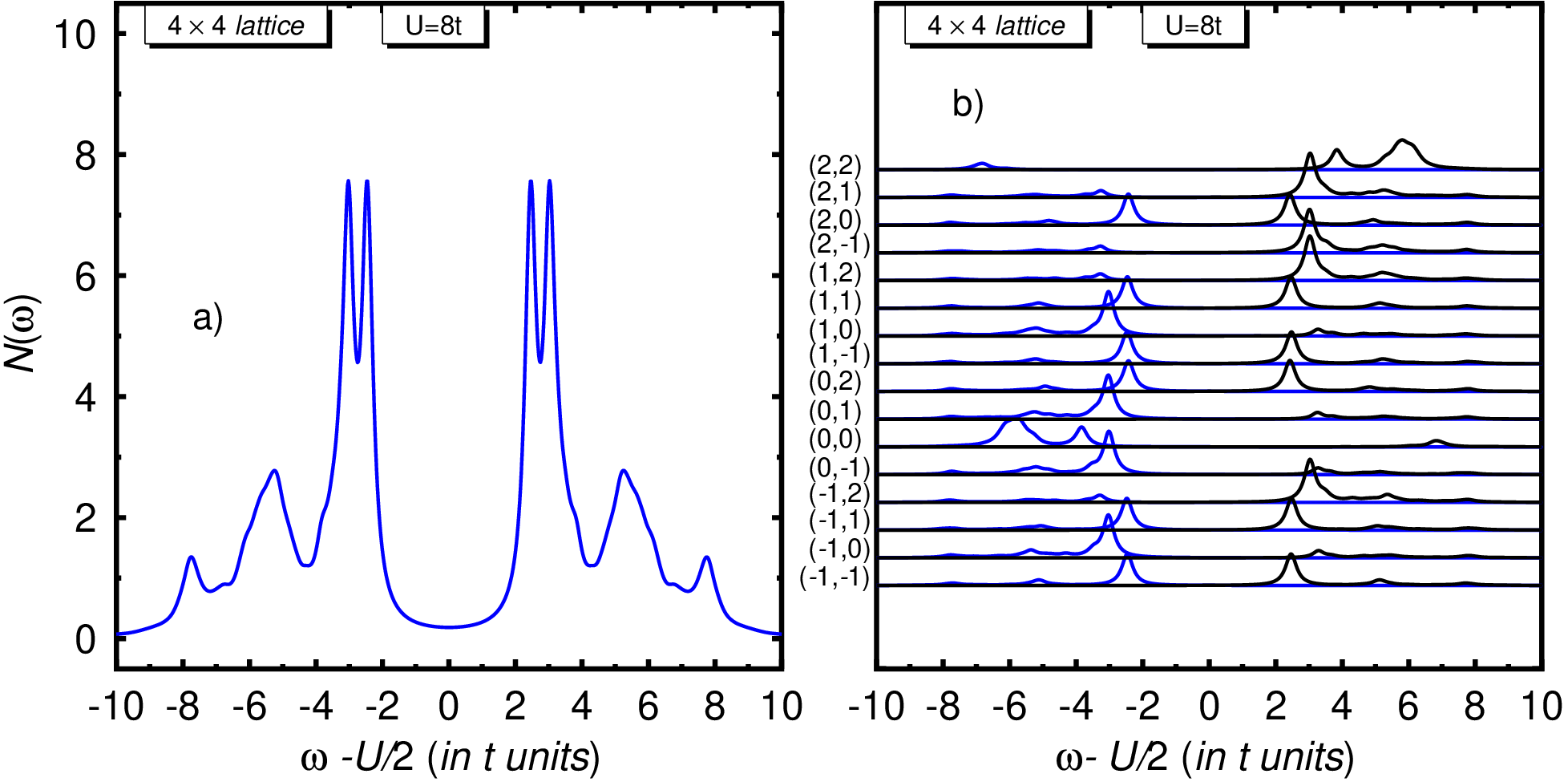} 
\caption{(Color online) The same as Fig.\ref{densitystates_halfilling_4by4_U4}
but for U=8t. The shapes of the DOS as well as the spectral functions 
 for momenta $(\pi,0)$, $(\pi/2,0)$ and $(0,0)$
 are qualitatively similar to the ones 
 obtained using Lanczos calculations. \cite{Dagotto-RC-3184}  
}
\label{densitystates_halfilling_4by4_U8} 
\end{figure*} 
%%%%%%%%%%%%%%%%%%%%%%%%%%%%%%%%%%%%%%%%%%%%%%%%%%%%%%%%%%%%%%%%%%%%%%%%%%%%%%%%%%%%%%%%%%%%%%%%%%%%
%
%  FIGURE 6 OF THE PAPER
%
%%%%%%%%%%%%%%%%%%%%%%%%%%%%%%%%%%%%%%%%%%%%%%%%%%%%%%%%%%%%%%%%%%%%%%%%%%%%%%%%%%%%%%%%%%%%%%%%%%% 

\begin{align} \label{hole-spectra--function}
&\langle h_{1}; {\Theta}^{-} || 
 {\hat{c}}_{{\boldsymbol{\xi}}^{0}-{\boldsymbol{\xi}}^{-}} || {\cal{D}}^{1}; {\Theta}^{0} \rangle
= 
-
\frac{1}{8 {\pi}^{2} N_{sites}}
\sqrt{\frac{2}{\langle {\cal{D}}^{1} |
\hat{P}_{00}^{{\Theta}^{0}} | {\cal{D}}^{1} \rangle}}
\times
\nonumber\\
&
\sum_{ih h^{'}{\sigma} {\sigma}^{'}}
f_{i h {\sigma}, h_{1}}^{{\Theta}^{-} *} 
\sum_{\boldsymbol{j}}
e^{-i {\bf{k}}_{\xi} {\bf{j}}}
\int d\Omega D_{\sigma {\sigma}^{'}}^{1/2 *}(\Omega)
(-1)^{1/2- {\sigma}^{'}}
\times
\nonumber\\ 
&
{\cal{D}}^{1 *}_{{\boldsymbol{\xi}}^{0}-{\boldsymbol{\xi}}^{-} -{\sigma}^{'}, h^{'}}
\Big[{\cal{X}}_{h^{'} h}^{i1} (\Omega,{\boldsymbol{j}}) \Big]^{-1} 
n^{i 1}(\Omega,{\boldsymbol{j}}) 
\end{align} 
where  ${\bf{k}}_{\xi}= \left(k_{{\xi}_{x}^{-}},k_{{\xi}_{y}^{-}}\right)
=
\left(\frac{2\pi {\xi}_{x}^{-}}{N_{x}},\frac{2\pi {\xi}_{y}^{-}}{N_{y}} \right)$. The 
indices i, $h,h^{'}$ and $\sigma, {\sigma}^{'}$ run 
as in (\ref{hole-wf-spectral}), ${\xi}_{x}^{-}$ 
and ${\xi}_{y}^{-}$ run as in Eq.(\ref{values-momenta}) and $\delta \epsilon_{h_{1}}= E^{{\Theta}^{0}}- E_{h_{1}}^{{\Theta}^{-}}$.  Details for 
the computation of the kernels ${\cal{H}}^{{\Theta}^{-}}$ and ${\cal{N}}^{{\Theta}^{-}}$ in Eq.(\ref{HW-holes-H})
as well as $\Big[{\cal{X}}_{h^{'} h}^{i1} (\Omega,{\boldsymbol{j}}) \Big]^{-1}$  
and $n^{i 1}(\Omega,{\boldsymbol{j}})$ in Eq.(\ref{hole-spectra--function}) can be found in appendices \ref{App-3} and \ref{App-1}, respectively. The occupation number
$n({\boldsymbol{\xi}}^{-})$
of a basis
state (\ref{Fourier-2D}) in the $N_{e}$-electron ground state 
can be computed as

\begin{eqnarray} \label{occupation-numbers}
\sum_{h_{1}=1}^{2n_{T}N_{e}}  
S_{h_{1}}({\boldsymbol{\xi}}^{-}, \delta \epsilon_{h_{1}})
= n({\boldsymbol{\xi}}^{-})
\end{eqnarray}

The ($N_{e}$+1)-electron system, with the symmetry
${\Theta}^{+}= (S=1/2,{\boldsymbol{\xi}}^{+})$, is approximated by  \cite{Carlos-Hubbard-1D}

\begin{eqnarray} \label{particle-wf-spectral}
|p_{1}; {\Theta}^{+}; \sigma  \rangle &=& 
\sum_{i p {\sigma}^{'}}
g_{i p {\sigma}^{'},p_{1}}^{{\Theta}^{+}}
\hat{P}^{{\Theta}^{+}}_{ \sigma {\sigma}^{'}}
\hat{b}^{\dagger}_{p}({\cal{D}}^{i}) | {\cal{D}}^{i} \rangle 
\end{eqnarray}
where the index i runs again as in (\ref{hole-wf-spectral}). The  particle index p 
takes the values
 $p=N_{e}+1, \dots, 2N_{sites}$ 
and ${\sigma}^{'}= \pm 1/2$. In this case, the coefficients $g^{{\Theta}^{+}}$ 
are obtained by solving the equation

\begin{eqnarray} \label{HW-particles-H}
\left( {\cal{H}}^{{\Theta}^{+}} -
 E_{p_{1}}^{{\Theta}^{+}}
 {\cal{N}}^{{\Theta}^{+}} \right)
 g^{{\Theta}^{+}} = 0
\end{eqnarray}
that yields 
$2n_{T}(2N_{sites}-N_{e})$ particle 
 solutions $p_{1}$ with
 energies $E_{p_{1}}^{{\Theta}^{+}}$.
The particle spectral function is then written as
$
S_{p_{1}}({\boldsymbol{\xi}}^{+}, \delta \epsilon_{p_{1}})
=
|\langle p_{1}; {\Theta}^{+} || 
{\hat{c}}_{{\boldsymbol{\xi}}^{+}-{\boldsymbol{\xi}}^{0}} || {\cal{D}}^{1}; {\Theta}^{0} \rangle|^{2}$
in terms of the reduced matrix element

\begin{align} \label{particle-spectra--function}
& 
\langle p_{1}; {\Theta}^{+} || 
{\hat{c}}_{{\boldsymbol{\xi}}^{+}-{\boldsymbol{\xi}}^{0}} || {\cal{D}}^{1}; {\Theta}^{0} \rangle
= 
-
\frac{1}{8 {\pi}^{2} N_{sites}}
\sqrt{\frac{2}{\langle {\cal{D}}^{1} |
\hat{P}_{00}^{{\Theta}^{0}}
| {\cal{D}}^{1} \rangle}}
\times
\nonumber\\
&
\sum_{ipp^{'}{\sigma} {\sigma}^{'}}
g_{i p \sigma,p_{1}}^{{\Theta}^{+} *} 
\sum_{\boldsymbol{j}}
e^{-i {\bf{k}}_{\xi} {\bf{j}}}
\int d\Omega D_{\sigma {\sigma}^{'}}^{1/2 *}(\Omega) 
n_{pp^{'}}^{i 1}(\Omega,{\boldsymbol{j}})
\times
\nonumber\\ 
&
{\cal{D}}^{1 *}_{{\boldsymbol{\xi}}^{+}-{\boldsymbol{\xi}}^{0} {\sigma}^{'}, p^{'}}
n^{i 1}(\Omega,{\boldsymbol{j}})
\end{align} 
where, in this case, ${\bf{k}}_{\xi}= \left(k_{{\xi}_{x}^{+}},k_{{\xi}_{y}^{+}}\right)
=
\left(\frac{2\pi {\xi}_{x}^{+}}{N_{x}},\frac{2\pi {\xi}_{y}^{+}}{N_{y}} \right)$.
The indices i, $p,p^{'}$ and $\sigma$, ${\sigma}^{'}$ run 
as in (\ref{particle-wf-spectral}), ${\xi}_{x}^{+}$ 
and ${\xi}_{y}^{+}$ run as in Eq.(\ref{values-momenta}) 
and $\delta \epsilon_{p_{1}}= E_{p_{1}}^{{\Theta}^{+}} - E^{{\Theta}^{0}}$.
Details for the computation of the kernels ${\cal{H}}^{{\Theta}^{+}}$
and ${\cal{N}}^{{\Theta}^{+}}$ 
as well as $n_{pp^{'}}^{i 1}(\Omega,{\boldsymbol{j}})$ 
 in Eq.(\ref{particle-spectra--function}) can be found in appendix \ref{App-3}.

Finally, the DOS can  be 
computed as

\begin{eqnarray} \label{densityofstates}
{\cal{N}}(\omega) = 
\sum_{{\boldsymbol{\xi}}}
\Big[
S_{(h_{1})}({\boldsymbol{\xi}},\omega) +
S_{(p_{1})}({\boldsymbol{\xi}},\omega)
\Big]
\end{eqnarray}
where the indices $h_{1}$ and $p_{1}$ are absorbed into the continuous 
variable $\omega$. Due to the finite size of the system the spectral functions 
consist of a finite number of $\delta$ functions with different weights. Therefore, we
introduce an artificial width $\Gamma$ for each state using a Lorentzian. In all cases 
our DOS is normalized to 2 $\times$ $N_{sites}$.

%%%%%%%%%%%%%%%%%%%%%%%%%%%%%%%%%%%%%%%%%%%%%%%%%%%%%%%%%%%%%%%%%%%%%%%%%%%%%%%%%%%%%%%%%%%%%%%%%%%%%%%%
%
% FIGURE 7 OF THE PAPER
%
%%%%%%%%%%%%%%%%%%%%%%%%%%%%%%%%%%%%%%%%%%%%%%%%%%%%%%%%%%%%%%%%%%%%%%%%%%%%%%%%%%%%%%%%%%%%%%%%%%%%%%%
\begin{figure*} 
\includegraphics[width=0.80\textwidth]{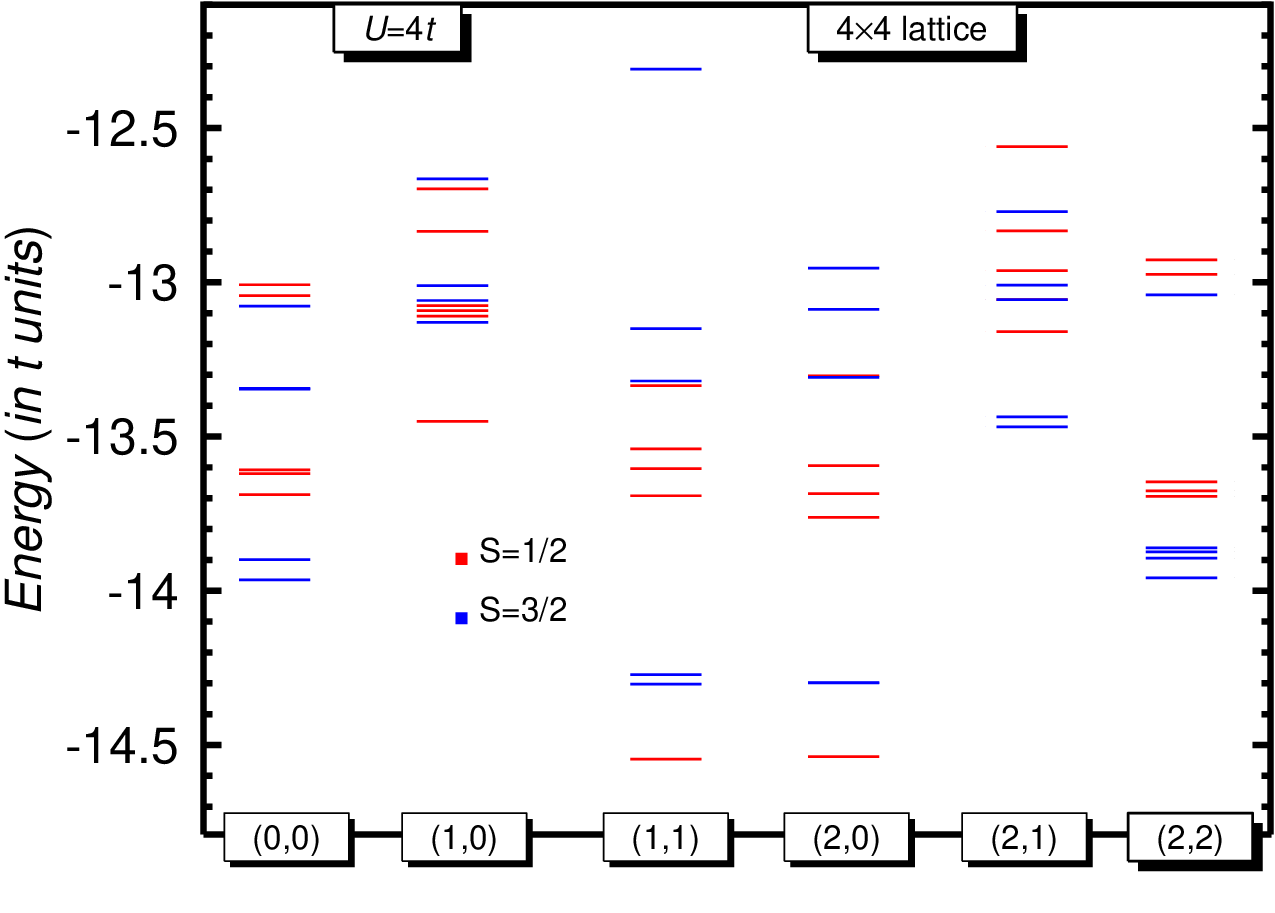} 
\caption{(Color online) Energy spectrum, obtained via Eq.(\ref{diagonalizationfinal}),
for the $4 \times 4$ lattice with $N_{e}=15$ electrons at U=4t.
}
\label{4by4_lattice_spectrum_Ne15}
\end{figure*}
%%%%%%%%%%%%%%%%%%%%%%%%%%%%%%%%%%%%%%%%%%%%%%%%%%%%%%%%%%%%%%%%%%%%%%%%%%%%%%%%%%%%%%%%%%%%%%%%%%%%%%%%
%
% END OF FIGURE 7 OF THE PAPER
%
%%%%%%%%%%%%%%%%%%%%%%%%%%%%%%%%%%%%%%%%%%%%%%%%%%%%%%%%%%%%%%%%%%%%%%%%%%%%%%%%%%%%%%%%%%%%%%%%%%%%%%%

\section{Discussion of results}
\label{results}
In this section, we discuss the results of our study. We have considered the $2 \times4$ half-filled
lattice as a prototypical system where one can obtain the full spectrum by means of  ED. This allows us to callibrate our approximation not only for ground state properties 
but also for excited states. Next, we have considered the well-studied half-filled $4 \times 4$ lattice, 
which constitutes the largest square lattice for which exact ground state energies 
are available in the literature. Other approximation
schemes have also been tested for this lattice in previous works. Results have already been  published for 
doped systems with 14 and 15 electrons in this lattice, which motivated us to also perform calculations 
for them in the present study. Last, we consider the half-filled $6 \times 6$ lattice as a prototype
of a system where  ED is no longer feasible. Many of the results
to be discussed in what follows correspond to 
U=4t taken as a representative on-site repulsion for which studies are 
available. Nevertheless, let us stress that  our formalism can be used 
for any 2D Hubbard hamiltonian of the form (\ref{HAM-HUB-2D})
with arbitrary U and/or t values.

\subsection{The square $2 \times 4$ lattice} 
\label{2by4results}
 
Let us start by  considering
 the rectangular $2 \times 4$ lattice. The first five solutions
  obtained
at half-filling
 via Eq.(\ref{diagonalizationfinal}),  for each of 
the linear momentum quantum numbers (0,0), (0,1), (0,2), (0,3), (1,0), (1,1), (1,2) and (1,3)
and the spins S=0,1, and 2, are plotted in panel a) of Fig.\ref{2by4_lattice_spectrum_Ne8} 
for U=4t. The first excited state 
corresponds to a 
 $\Theta=(1,1,2)$  
 configuration [with linear 
 momenta $(\pi,\pi)$]. The energies 
 ${\epsilon}_{\alpha}^{\Theta}$ 
 of the 120 solutions shown in the figure, have been compared to the ones
 obtained using an ED. \cite{Albuquerque} The comparison reveals that  
 both spectra follow the same qualitative trend and 
 can hardly be distinguished. Therefore in panel b) of the same figure, we have plotted the absolute errors 
$e_{absol.}=E_{exact}-E$ for each of the predicted 120 states. Our 
approximation 
fairly reproduces the exact ground state energy -10.2529t 
for this system. For all the 40 S=0 and S=1 solutions considered 
the absolute errors remain very small, the largest deviation being 0.047t
for the second state with the symmetry $\Theta=(1,0,0)$. The previous results are encouraging 
if one takes into account that, even for this relatively small lattice, the number of variational parameters 
in our approximation
$n_{var}$($S=0,{\xi}_{x},{\xi}_{y}) = 128$   and $n_{var}$($S=1,{\xi}_{x},{\xi}_{y}) = 132$ is 
about half of the dimensions 
$n_{RH}$($S=0,{\xi}_{x},{\xi}_{y}) = 221$ and  $n_{RH}$($S=1,{\xi}_{x},{\xi}_{y}) = 294$
of the restricted Hilbert spaces. On the other hand, $n_{var}$($S=2,{\xi}_{x},{\xi}_{y}) = 136$ is larger than 
 $n_{RH}$($S=2,{\xi}_{x},{\xi}_{y})=90$ and therefore our solutions reproduce the ED ones for S=2 states.

In panel a) of Fig. \ref{densitystates_halfilling_2by4_U4}, we have plotted the DOS 
${\cal{N}}(\omega)$ [Eq.(\ref{densityofstates})]
for the 
half-filled $2 \times 4$ lattice at U=4t. The calculations have been carried out by 
approximating the  ($N_{e}$ $\pm$ 1)-electron systems 
[see Eqs. (\ref{hole-wf-spectral}) and (\ref{particle-wf-spectral})] 
with $n_{T}$=1 (red curve) and $n_{T}$=5 (blue curve) HF-transformations  
along the lines described in Sec. \ref{spectral-H2D}.
We have introduced a shift equal to the chemical potential at 
half-filling (${\mu}_0 = U/2$) so that the DOS in Fig.\ref{densitystates_halfilling_2by4_U4}  appears to be
symmetric around $\omega$-U/2=0. This convention, i.e., to plot DOS 
and spectral functions vs.  $\omega$-U/2 will be adopted in the rest of the paper.
The DOS shows the
Hubbard gap, $\Delta_H = U/2 = 2t$, characteristic of finite size
lattices. We note, however, that previous studies 
within the framework of the dynamical cluster approximation (DCA)
have shown that the gap is preserved at sufficiently low 
temperatures even in the thermodynamic limit (TDL). \cite{moukouri2001, huscroft2001,aryanpour2003}
On the other hand, the nonperturbative study of Ref. 59  has  concluded
that for the half-filled Hubbard model the gap persists for any finite value
of the on-site repulsion U, the only singular point being U=0t.

%%%%%%%%%%%%%%%%%%%%%%%%%%%%%%%%%%%%%%%%%%%%%%%%%%%%%%%%%%%%%%%%%%%%%%%%%%%%%%%%%%%%%%%%%%%%%%%%%%%%%%%%%%%%
%
%  FIGURE 8 OF THE PAPER
%
%%%%%%%%%%%%%%%%%%%%%%%%%%%%%%%%%%%%%%%%%%%%%%%%%%%%%%%%%%%%%%%%%%%%%%%%%%%%%%%%%%%%%%%%%%%%%%%%%%%%%%%%%%%%
\begin{figure*} 
\includegraphics[width=0.70\textwidth]{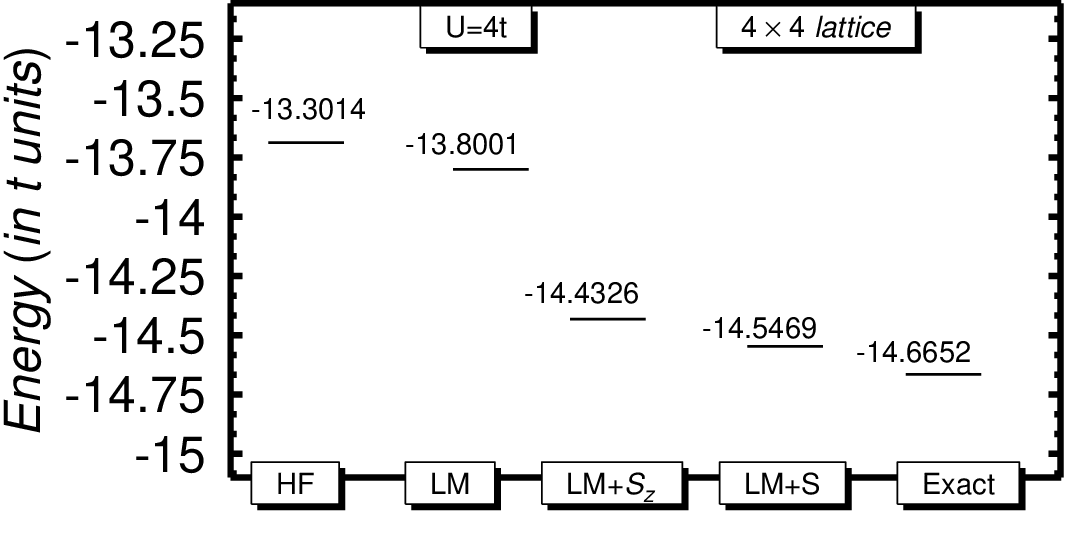} 
\caption{ Ground state energy of the $4 \times 4$ lattice with $N_{e}=15$
electrons at U=4t computed with various approaches. The different columns refer to the 
unprojected Hartree-Fock (HF) calculation, HF with linear momentum projection (LM), HF 
with projection of linear momentum and only the z-component of the total spin  (LM+S$_{z}$)
and HF with projection of linear momentum and full spin projection before the variation
(LM+S). For all these methods we have used the approximation discussed in Sec. \ref{formalism-H2D-excited}
with five transformations. Note that the LM+S method corresponds to the symmetry-projected 
configuration mixing approach used 
throughout the paper. The predicted energies are compared with the exact (EXACT) 
one. \cite{Lanczos-Fano}
For details, see the main text.
}
\label{oneholeprogress}
\end{figure*}
%%%%%%%%%%%%%%%%%%%%%%%%%%%%%%%%%%%%%%%%%%%%%%%%%%%%%%%%%%%%%%%%%%%%%%%%%%%%%%%%%%%%%%%%%%%%%%%%%%%%%%%%%%%%
%
%  END OF FIGURE 8 OF THE PAPER
%
%%%%%%%%%%%%%%%%%%%%%%%%%%%%%%%%%%%%%%%%%%%%%%%%%%%%%%%%%%%%%%%%%%%%%%%%%%%%%%%%%%%%%%%%%%%%%%%%%%%%%%%%%%%%

From panel a) of Fig.\ref{densitystates_halfilling_2by4_U4} one 
realizes that, even for this small lattice, the fine details 
of the energy distribution of ${\cal{N}}(\omega)$ can only be obtained using a larger number 
 $n_{T}$=5
 of HF-transformations
to describe the  ($N_{e}$ $\pm$ 1)-electron systems. Using  $n_{T}$=5 transformations, Eqs.(\ref{HW-holes-H}) and (\ref{HW-particles-H})
provide us with 80 hole and particle solutions while only 16 solutions are obtained with   $n_{T}$=1. 
Therefore, contributions  to ${\cal{N}}(\omega)$ with a more collective nature can be better accounted for in the former 
case (i.e.,  $n_{T}$=5).  This is further corroborated by comparing our DOS, computed with $n_{T}$=5  transformations, with 
the one obtained using an ED, performed with an in-house code, shown in panel b) of the figure. 
Note that we have intentionally used a small broadening $\Gamma$=0.05t to 
retain as much structure as possible in our 
DOS as well as to emphasize the differences with the ED one. As can be observed
there is excellent agreement in the position and relative heights of all the prominent peaks.
The
hole (blue) and particle (black)
spectral functions, computed with  $n_{T}$=5 HF-transformations, are displayed in panel c) of the 
same figure. We have not included the ones provided by the ED since they are quite similar to ours.
Their structure  is dominated by a main peak 
but less prominent ones are also visible in the figure. 
The momenta $(0,\pi)$ and $(\pi,0)$, at the noninteracting
Fermi surface $\epsilon({\bf{k}}_{\alpha})=0$ [see, Eq.(\ref{eq-seenergies}) of appendix \ref{App-1}], have the 
largest spectral weight near $\omega$-U/2=0. On the other hand, the
momenta $(0,0)$ and $(0,\pm \pi/2)$ [$(\pi, \pm \pi/2)$ and $(\pi,\pi)$]  inside 
(outside)
the noninteracting Fermi surface 
contribute mostly to hole (particle) states.

In Fig. \ref{occupations_halfilling_2by4_U4}, we display the occupation numbers of  the basis states [see Eq.(\ref{Fourier-2D})] 
in the ${\Theta}^{0}=(0,0,0)$ ground state of the half-filled $2 \times 4$ lattice. Results are 
shown for the on-site
repulsions U=4t, 20t, 40t, 64t, 80t and 120t. The calculations were performed  using 80 hole solutions ${\overline{h}}$ 
(i.e.,  $n_{T}$=5 HF-transformations) in Eq.(\ref{occupation-numbers}). The evolution of the occupations clearly depict the 
transition to the strong coupling regime where 
the Hubbard Hamiltonian \cite{Hubbard-model_def1}  can 
be mapped into the AF Heisenberg model. \cite{text-Hubbard-1D}
 In fact, for U $\ge$ 64t  the results 
look very similar 
to the  uniform  distribution, with occupations $n({\boldsymbol{\xi}^{-}})=1$, expected in the   limit 
 U $ \rightarrow \infty$.

\subsection{The square $4 \times 4$ lattice} 
\label{4by4results}

In panel a) of Fig.\ref{4by4_lattice_spectrum_Ne16}, we show  the energies
${\epsilon}_{\alpha}^{\Theta}$ 
obtained, via Eq.(\ref{diagonalizationfinal}), for 
the half-filled $4 \times 4$ lattice at U=4t. Results are only shown for the six essentially different 
pairs of linear momentum quantum numbers (0,0), (1,0), (1,1), (2,0), (2,1) and (2,2).  For 
each of them we have plotted the energies of 
the first five solutions with spins S=0,1, and 2. In this case, the number of variational 
parameters 
in our approximation
is $n_{var}$($S=0,{\xi}_{x},{\xi}_{y}$)=512,  $n_{var}$($S=1,{\xi}_{x},{\xi}_{y}$)=516
and $n_{var}$($S=2,{\xi}_{x},{\xi}_{y}$)=520 while the dimensions of the restricted Hilbert spaces 
are $n_{RH}(S=0,{\xi}_{x},{\xi}_{y}) \approx 2 \times 10^{6}$, $n_{RH}(S=1,{\xi}_{x},{\xi}_{y}) \approx 4 \times 10^{6}$
and $n_{RH}(S=2,{\xi}_{x},{\xi}_{y}) \approx 3 \times 10^{6}$, respectively.

The energy -13.5898t of our 
${\Theta}^{0}=(0,0,0)$  ground state accounts for 
99.76 $\%$ of the exact one, \cite{Neuscamman-2012,Lanczos-Fano} -13.6219t. 
In order to put our result in perspective, the relative 
error 0.24 $\%$ in our ground state energy per site  
${\epsilon}_{1}^{{\Theta}^{0}}/16$
at U=4t
should  be compared, for example, with the value
0.70 $\%$ recently
reported \cite{Neuscamman-2012} within the framework of the 
variational MC (VMC) approximation using an  ansatz, 
consisting of 
the product of a
correlator product
state tensor network
 and a Pfaffian wave function, with 524,784 variational parameters. Note, that
  the DMRG formalism in momentum space \cite{Xiang} (kDMRG) predicts a relative 
 error of 0.37$\%$. We have also studied  
 our ground state energy per site 
 ${\epsilon}_{1}^{{\Theta}^{0}}/16$
 as a function of the interaction strength
 U=2t, 6t, 8t, 10t, 12t and 16t and found relative errors
 always smaller than 0.4 $\%$.

%%%%%%%%%%%%%%%%%%%%%%%%%%%%%%%%%%%%%%%%%%%%%%%%%%%%%%%%%%%%%%%%%%%%%%%%%%%%%%%%%%%%%%%%%%%%%%%%%%%%%%%%%%%%
%
% FIGURE 9 OF THE PAPER
%
%%%%%%%%%%%%%%%%%%%%%%%%%%%%%%%%%%%%%%%%%%%%%%%%%%%%%%%%%%%%%%%%%%%%%%%%%%%%%%%%%%%%%%%%%%%%%%%%%%%%%%%%%%%%
\begin{figure*} 
\includegraphics[width=1.0\textwidth]{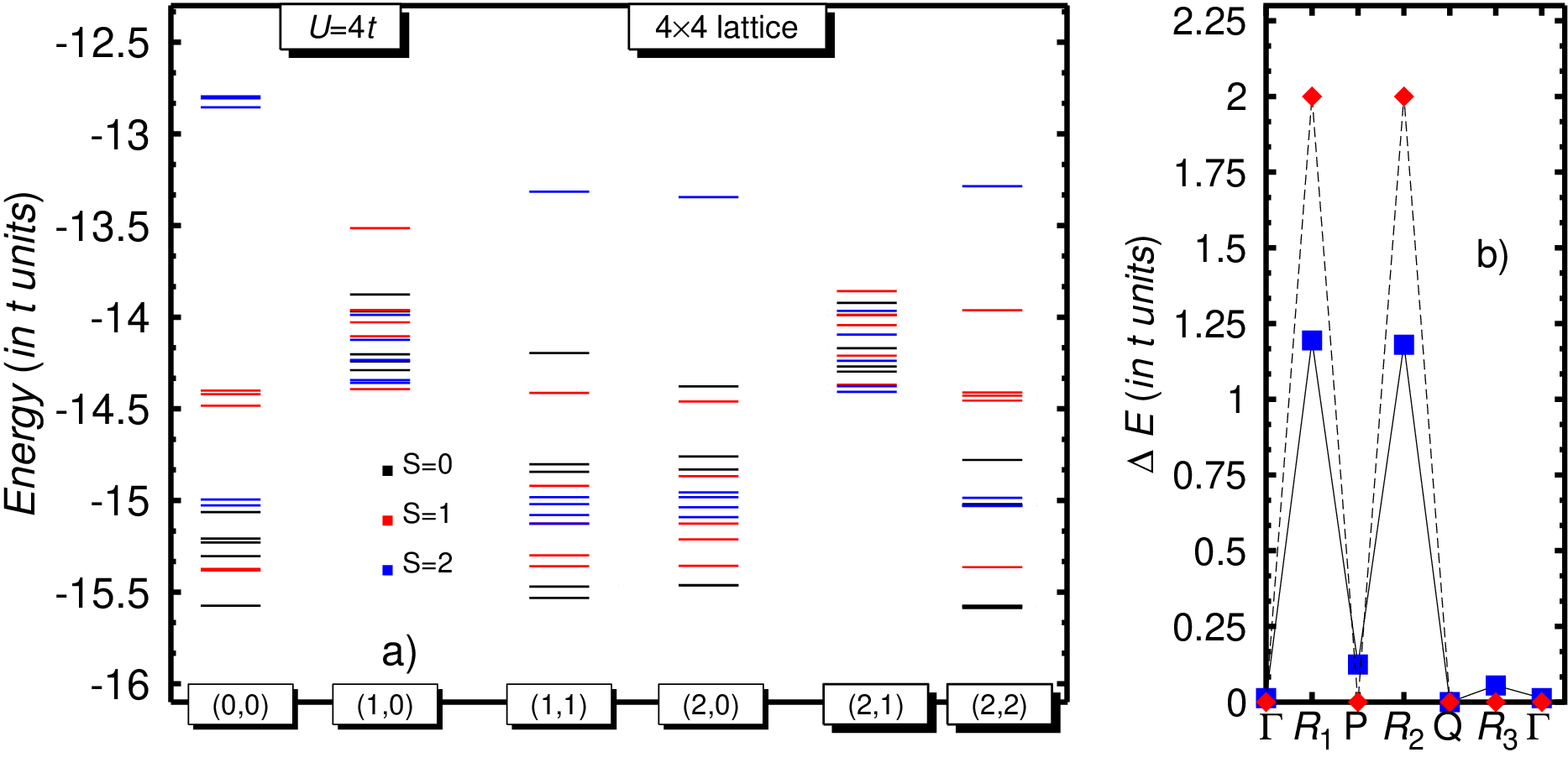} 
\caption{(Color online) The energy spectrum,  obtained via Eq.(\ref{diagonalizationfinal}), for
 the  $4 \times 4$ lattice with $N_{e}=14$ electrons at U=4t is shown in panel a). 
In panel b), the excitation energies 
from the ground state to the lowest-lying S=0,1, and S=2 states 
from panel a) are plotted as functions 
 of the linear momentum quantum numbers $\Gamma=(0,0)$, $R_{1}=(1,0)$, $P=(2,0)$, $R_{2}=(2,1)$, $Q=(2,2)$, and 
$R_{3}=(1,1)$, respectively. In addition to U=4t (blue boxes), results for U=0t (red diamonds) are also included for comparison.  
}
\label{4by4_lattice_spectrum_Ne14}
\end{figure*} 
%%%%%%%%%%%%%%%%%%%%%%%%%%%%%%%%%%%%%%%%%%%%%%%%%%%%%%%%%%%%%%%%%%%%%%%%%%%%%%%%%%%%%%%%%%%%%%%%%%%%%%%%%%%%
%
% FIGURE 9 OF THE PAPER
%
%%%%%%%%%%%%%%%%%%%%%%%%%%%%%%%%%%%%%%%%%%%%%%%%%%%%%%%%%%%%%%%%%%%%%%%%%%%%%%%%%%%%%%%%%%%%%%%%%%%%%%%%%%%% 

Coming back to the spectrum 
shown in panel a) of 
Fig.\ref{4by4_lattice_spectrum_Ne16}, we note that the first excited state 
corresponds to a 
 $\Theta=(1,2,2)$  
 configuration [with linear 
 momenta $(\pi,\pi)$]. In fact, similar to the 
 half-filled $2 \times 4$ lattice, the 
 first excited state for each combination  $({\xi}_{x},{\xi}_{y})$ 
 has spin S=1, exception made of $(0,0)$. The 
 $2 \times 4$ lattice displays a low-lying S=0 singlet   
 (see Fig.\ref{2by4_lattice_spectrum_Ne8}) while 
 an S=2 quintet appears in the $4 \times 4$ lattice. The excitation energies, referred 
 to the ${\Theta}^{0}=(0,0,0)$  configuration, of these
 low-lying S=1 and S=2  states are shown in
 panel b) of Fig.\ref{4by4_lattice_spectrum_Ne16} as functions of the linear momentum quantum numbers. 
 The shape of the curve does not fully agree with 
 the one obtained with the
 spin-density wave (SDW) 
 approximation \cite{Fawcett-sdw} mainly due to the absence of degeneracy between 
 the $\Gamma$=(0,0) and $Q$=(2,2)  as well as the two peaks for the 
 $R_{1}$=(1,0) and $R_{2}$=(2,1)
 points. Much of this discrepancy could, however, be due to finite size effects.  \cite{Lanczos-Fano} Note that 
 the two peaks at  $R_{1}$ and $R_{2}$, resulting from a kinetic-energy gap of 2t, are
  already visible for the Fermi gas (U=0t).

The DOS 
${\cal{N}}(\omega)$ [Eq.(\ref{densityofstates})]
for the 
half-filled $4 \times 4$ lattice 
at U=4t
is shown in panel a) of Fig.\ref{densitystates_halfilling_4by4_U4}.    
The calculations 
have been performed 
using  $n_{T}$=5   HF-transformations
along the lines described in Sec. \ref{spectral-H2D}. 
In 
this case, Eqs.(\ref{HW-holes-H}) and (\ref{HW-particles-H}) provide us with 
160 hole and particle solutions. 
A Lorentzian folding of width  $\Gamma$=0.2t has been used. Similar to the case 
of the half-filled $2 \times 4$ lattice, the 
Hubbard gap, $\Delta_H = U/2 =2t$, remains present in this larger system.
The hole (blue) and particle (black) spectral
functions shown in panel b) of the figure show that the momenta 
$(\pm \pi/2,\pm \pi/2)$, $(0,\pi)$ and $(\pi,0)$ at the noninteracting 
Fermi surface have
the largest spectral weight near $\omega$-U/2=0. Moreover, the
 spectral weight due to these momenta at the Fermi
surface is particle-hole symmetric.

In panel a) of Fig.\ref{densitystates_halfilling_4by4_U8}, we have plotted the DOS 
${\cal{N}}(\omega)$ [Eq.(\ref{densityofstates})]
for the half-filled $4 \times 4$
lattice at U=8t (i.e., an on-site repulsion equal to the noninteracting bandwidth W=8t)
computed with $n_{T}$=5   HF-transformations.
A Lorentzian folding of width  $\Gamma$=0.2t has been used.
The corresponding hole (blue) and particle (black) spectral functions 
are also  
displayed in panel b) of the figure.
Our DOS 
and spectral functions [in particular, the ones corresponding
to the linear momenta $(\pi,0)$, $(\pi/2,0)$ and $(0,0)$] 
 can be  compared with the ones, obtained using the Lanczos method, 
 shown in Fig.2  of Ref. 57. As can be observed, the main 
 qualitative
 features of the
particle-hole symmetric 
 DOS are well
reproduced, namely the two prominent peaks at $\omega$-U/2 $\approx$ 2t and 3t (-2t and -3t)  a lump peaked around      
$\omega$-U/2 $\approx$ 5t (-5t) and a smaller satellite peak in the neighborhood  of  $\omega$-U/2 $\approx$ 8t (-8t).
In agreement with  the results of Ref. 57, the upper and lower bands as well as the 
Hubbard gap are also clearly visible in Fig.\ref{densitystates_halfilling_4by4_U8}.

We have also studied 
the evolution of the 
DOS  for 
the half-filled $4 \times 4$  lattice as a function 
of U. To this end, in addition to the cases U=4t and 8t
shown in Figs. \ref{densitystates_halfilling_4by4_U4} and 
\ref{densitystates_halfilling_4by4_U8}, calculations have
also been performed for U=2t, 12t and 20t. In good agreement with 
previous studies, \cite{Dagotto-RC-3184,moukouri2001, huscroft2001,aryanpour2003,stanescu2001} we observe that the Hubbard 
gap persists for increasing values of U. In our 
calculations, a pronounced suppression in the DOS around $\omega$-U/2=0
is  observed with the DOS fully  vanishing around U=8t
(i.e., around the noninteracting bandwidth W=8t) which is precisely
the region where a sizeable Hubbard gap is developed. \cite{Dagotto-RC-3184,Zgid,stanescu2001}

%%%%%%%%%%%%%%%%%%%%%%%%%%%%%%%%%%%%%%%%%%%%%%%%%%%%%%%%%%%%%%%%%%%%%%%%%%%%%%%%%%%%%%%%%%%%%%%%%%%%%%%%%%%%
%
% FIGURE 10 OF THE PAPER
%
%%%%%%%%%%%%%%%%%%%%%%%%%%%%%%%%%%%%%%%%%%%%%%%%%%%%%%%%%%%%%%%%%%%%%%%%%%%%%%%%%%%%%%%%%%%%%%%%%%%%%%%%%%%%
\begin{figure*} 
\includegraphics[width=1.00\textwidth]{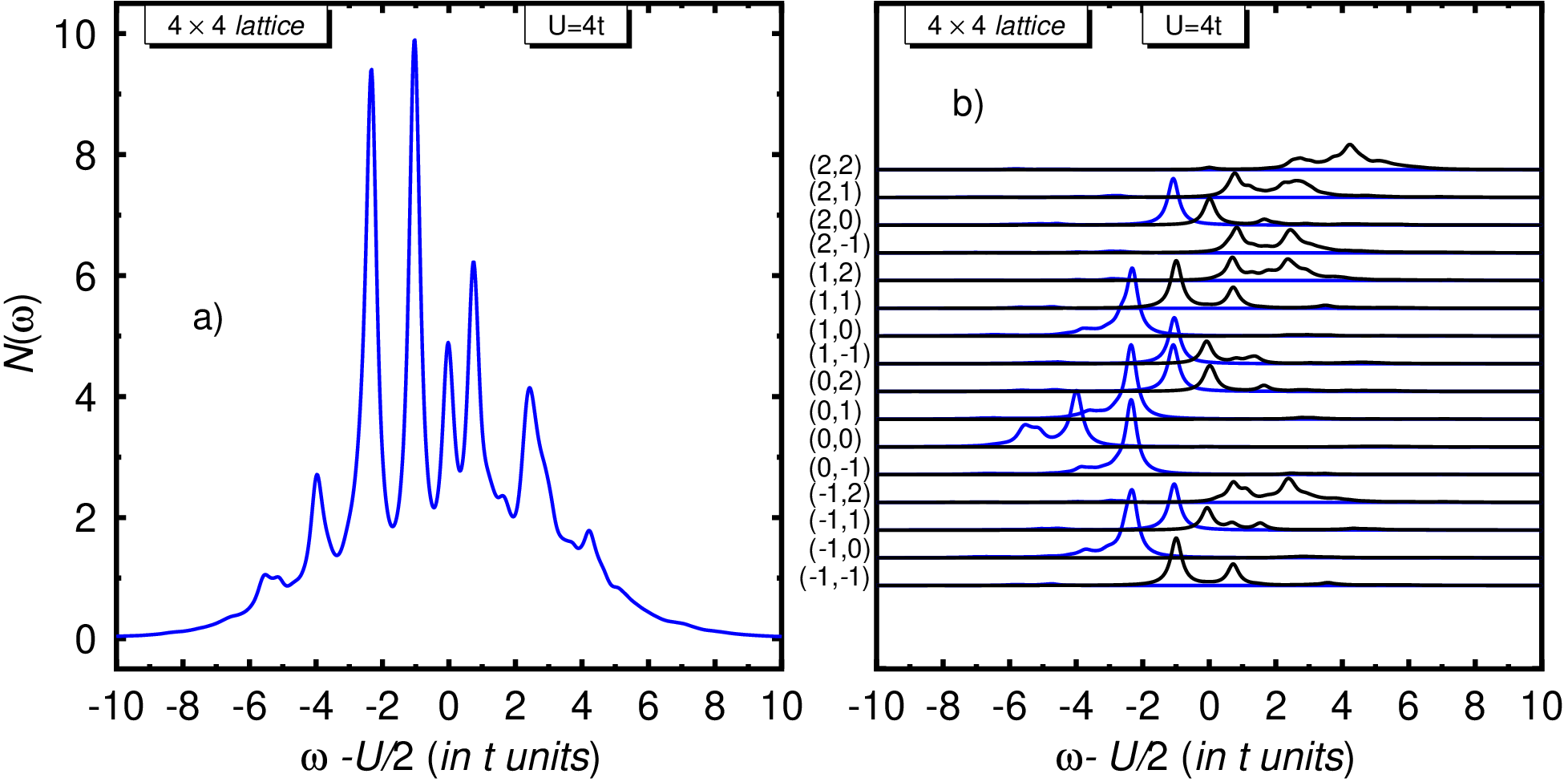} 
\caption{(Color online)
The same as Fig.\ref{densitystates_halfilling_4by4_U4} but for 
the  $4 \times 4$ lattice with $N_{e}$=14 electrons
at U=4t.
}
\label{densitystates_4by4_Ne14_U4} 
\end{figure*}
%%%%%%%%%%%%%%%%%%%%%%%%%%%%%%%%%%%%%%%%%%%%%%%%%%%%%%%%%%%%%%%%%%%%%%%%%%%%%%%%%%%%%%%%%%%%%%%%%%%%%%%%%%%%
%
% END FIGURE 10 OF THE PAPER
%
%%%%%%%%%%%%%%%%%%%%%%%%%%%%%%%%%%%%%%%%%%%%%%%%%%%%%%%%%%%%%%%%%%%%%%%%%%%%%%%%%%%%%%%%%%%%%%%%%%%%%%%%%%%%

Let us now consider two  examples of a doped $4 \times 4$ lattice. In 
Fig.\ref{4by4_lattice_spectrum_Ne15}, we show  the spectrum 
in the case of 15 electrons at U=4t. 
For each of the linear momentum quantum numbers (0,0), (1,0), (1,1), (2,0), (2,1) and 
(2,2), we plot the energies of the first five solutions of 
Eq.(\ref{diagonalizationfinal}) for the spins S=1/2 and 3/2. The number of variational parameters
in our approximation
$n_{var}$($S=1/2,{\xi}_{x},{\xi}_{y})$=512 
and $n_{var}$($S=3/2,{\xi}_{x},{\xi}_{y})$=516
should be compared with the dimensions  
$n_{RH}$($S=1/2,{\xi}_{x},{\xi}_{y}) \approx$ 2 $\times 10^{6}$
and $n_{RH}$($S=3/2,{\xi}_{x},{\xi}_{y}) \approx$ 2 $\times 10^{6}$ 
of the restricted Hilbert spaces. The first noticeable feature
in Fig.\ref{4by4_lattice_spectrum_Ne15} is that the four-fold degenerate 
${\Theta}^{-}=(1/2,1,1)$ ground state 
has non-zero linear momenta $(\pi/2,\pi/2)$. 
A finite linear momentum  for the one-hole
ground state has also been predicted in previous studies 
\cite{Dagotto-review} using a variety of  approximations for lattices of different sizes.
Our numerical calculations also
predict 
 a two-fold
degenerate (1/2,2,0) configuration whose energy is almost the same as the ground state one. 
For the noninteracting  system (U=0t), the lowest-lying 
S=1/2 and S=3/2
states 
with linear momentum quantum numbers  $\Gamma=(0,0)$, $P=(2,0)$, $Q=(2,2)$, and 
$R_{3}=(1,1)$ are degenerate and the same is also true for the configurations $R_{1}=(1,0)$ 
and  $R_{2}=(2,1)$. Therefore, the huge degeneracy observed in the noninteracting case 
is  already partially lifted at U=4t.

In Fig.\ref{oneholeprogress}, we compare the ground state energy
of the  $4 \times 4$ lattice with 15 electrons with the exact one \cite{Lanczos-Fano}
for U=4t. The energy -14.5469t
predicted within our symmetry-projected 
configuration mixing approach, via Eq.(\ref{diagonalizationfinal}), accounts 
for 99.19 $\%$ of the 
exact result. It is interesting to note that linear 
momentum plus $\hat{S}_{z}$ projection 
already accounts for 98.41 $\%$ of the exact solution. Nevertheless, full spin 
projection, while also recovering the total spin quantum number, still brings a 
sizeable amount 
of correlations.

The (shifted)
differences 
${\epsilon}_{1}^{{\Theta}^{0}}- {\epsilon}_{1}^{(1/2,{\xi}_{x},{\xi}_{y})}$-U/2, where 
${\epsilon}_{1}^{{\Theta}^{0}}$ is the ground state energy of the half-filled lattice 
(Fig.\ref{4by4_lattice_spectrum_Ne16})
and 
${\epsilon}_{1}^{(1/2,{\xi}_{x},{\xi}_{y})}$
represents the energy of each of the 
lowest-lying S=1/2 states in Fig.\ref{4by4_lattice_spectrum_Ne15}, compare very well 
with the  position of the first prominent peak  in the hole spectral functions shown in 
panel b) of Fig.\ref{densitystates_halfilling_4by4_U4}. For example, the variational 
approach predicts ${\epsilon}_{1}^{{\Theta}^{0}}- {\epsilon}_{1}^{(1/2,1,1)}$-U/2=-1.044t
while the corresponding peak 
in the hole spectral function is predicted 
to be at $\omega$-U/2= -1.010t. The same is also true for the configuration 
with linear momenta ($\pi$,0) for which the variational approach predicts 
${\epsilon}_{1}^{{\Theta}^{0}}- {\epsilon}_{1}^{(1/2,2,0)}$-U/2=-1.052t
whereas the position of the corresponding peak 
in the hole spectral function is predicted 
to be at $\omega$-U/2= -1.012t. This leads to the conclusion that
the lowest-lying S=1/2 states in the spectrum of Fig.\ref{4by4_lattice_spectrum_Ne15}
are reasonably well described by a wave function of the form (\ref{hole-wf-spectral}). This is remarkable, since 
no orbital relaxation is accounted for in this wave function, i.e., the determinants 
$| {\cal{D}}^{(i)} \rangle$ in Eq.(\ref{hole-wf-spectral})  correspond to the ones obtained 
at half-filling.

%%%%%%%%%%%%%%%%%%%%%%%%%%%%%%%%%%%%%%%%%%%%%%%%%%%%%%%%%%%%%%%%%%%%%%%%%%%%%%%%%%%%%%%%%%%%%%%%%%%%%%%%%%%
%
%  FIGURE 11 OF THE PAPER
%
%%%%%%%%%%%%%%%%%%%%%%%%%%%%%%%%%%%%%%%%%%%%%%%%%%%%%%%%%%%%%%%%%%%%%%%%%%%%%%%%%%%%%%%%%%%%%%%%%%%%%%%%%%
\begin{figure*} 
\includegraphics[width=1.00\textwidth]{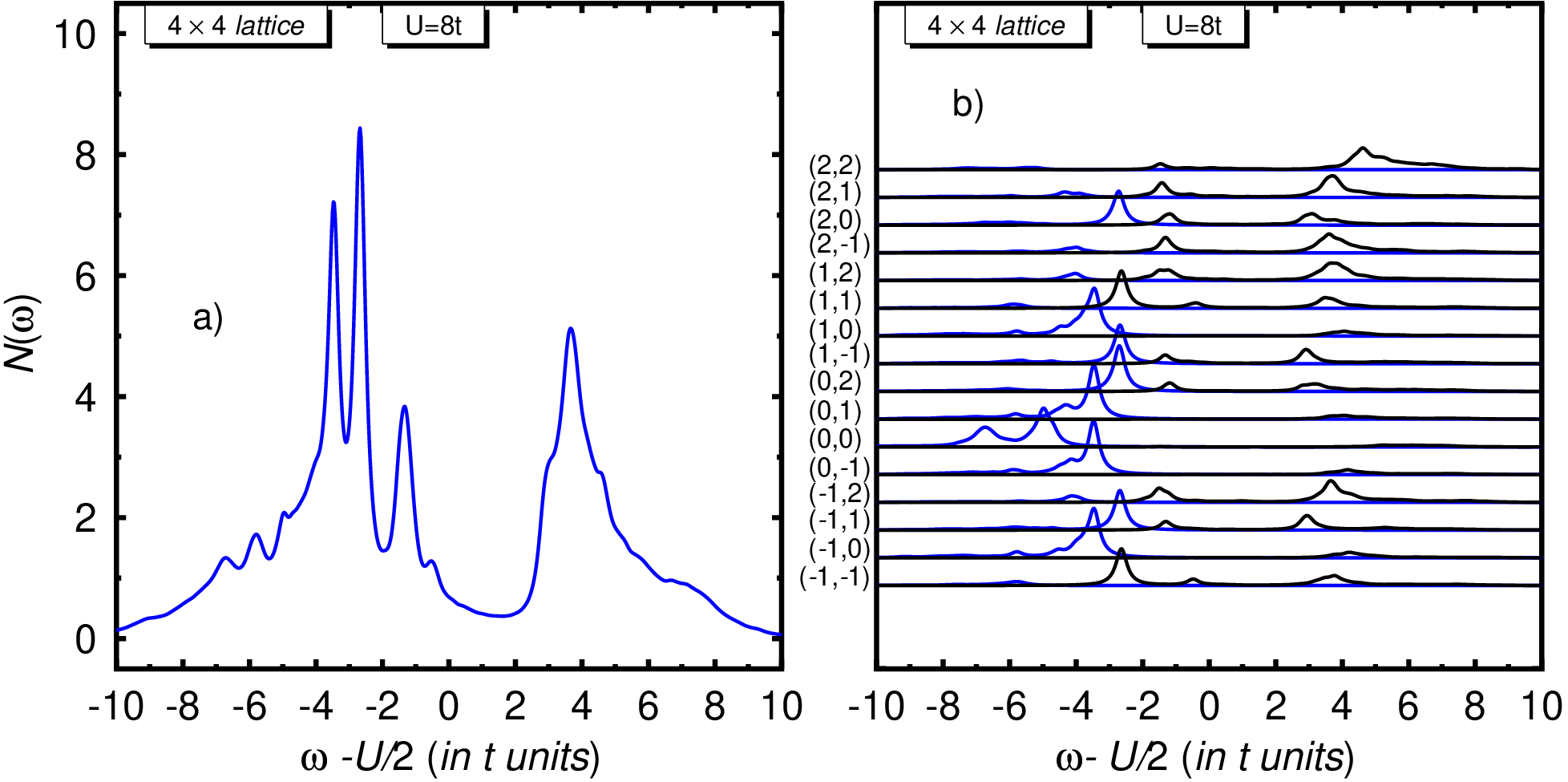} 
\caption{(Color online)
The same as Fig.\ref{densitystates_halfilling_4by4_U4} but for the  $4 \times 4$ lattice 
with $N_{e}$=14 electrons
at U=8t. The shapes of the DOS as well as the spectral functions 
 are qualitatively similar to the ones 
 obtained using Lanczos calculations. \cite{Dagotto-RC-3184} 
}
\label{densitystates_4by4_Ne14_U8} 
\end{figure*}
%%%%%%%%%%%%%%%%%%%%%%%%%%%%%%%%%%%%%%%%%%%%%%%%%%%%%%%%%%%%%%%%%%%%%%%%%%%%%%%%%%%%%%%%%%%%%%%%%%%%%%%%%%%
%
%  END OF FIGURE 11 OF THE PAPER
%
%%%%%%%%%%%%%%%%%%%%%%%%%%%%%%%%%%%%%%%%%%%%%%%%%%%%%%%%%%%%%%%%%%%%%%%%%%%%%%%%%%%%%%%%%%%%%%%%%%%%%%%%%%

The spectrum obtained,  via Eq.(\ref{diagonalizationfinal}), for 14 electrons at U=4t is displayed   
in Fig.\ref{4by4_lattice_spectrum_Ne14}. The 
number
of variational parameters
in our approximation
 $n_{var}$($S=0,{\xi}_{x},{\xi}_{y})$=504, $n_{var}$($S=1,{\xi}_{x},{\xi}_{y})$=508
and $n_{var}$($S=2,{\xi}_{x},{\xi}_{y})$=512
should be compared with the dimensions  
$n_{RH}$($S=0,{\xi}_{x},{\xi}_{y}) \approx 10^{6}$, $n_{RH}$($S=1,{\xi}_{x},{\xi}_{y}) \approx 2 \times 10^{6}$
and $n_{RH}$($S=2,{\xi}_{x},{\xi}_{y}) \approx 10^{6}$ of the restricted Hilbert spaces. The ground 
state corresponds to the ${\Theta}^{0}$=(0,2,2) configuration 
[linear momenta ($\pi,\pi)$]
with energy
-15.5872t, while the exact one is -15.7446t. \cite{Lanczos-Fano} On the other hand, the
VMC approximation \cite{Ericprivate} predicts a ground state energy of
-15.5936t. Thus, both methods, ours and VMC, yield essentially the same 
relative error of around 1 $\%$ in the ground state energy per site. On the 
other hand, a relative error in the ground state energy per site of 
0.45 $\%$ is obtained within the kDMRG approximation. \cite{Xiang}

Our calculations also predict
two other close-lying (0,2,2) solutions (with energies -15.5747t and -15.5777t) which cannot 
be distinguished in Fig.\ref{4by4_lattice_spectrum_Ne14} and therefore appear, together 
with the ground state, as a single thick black line. The energy
-15.5743t of the 
 ${\Theta}$=(0,0,0) configuration is also  close to the actual ground state. As a 
result, the symmetry of the $\Gamma$ and $Q$ points is almost recovered in 
panel b) of Fig.\ref{4by4_lattice_spectrum_Ne14}
where the energies of the
 lowest-lying S=0,1,2  states for each linear momentum combination are shown,  referred 
 to the  ${\Theta}^{0}=(0,2,2)$ ground state for this system. Note that the 
 configurations (0,1,1) and (0,2,0) (i.e., the points  $R_{3}$ and $P$ 
 in panel b) of Fig.\ref{4by4_lattice_spectrum_Ne14}) have very small excitation energies, 0.0552t and 
 0.1242t, respectively. The two peaks at the points  $R_{1}$ and $R_{2}$ 
 are 
 also present in the  system with $N_{e}=14$ electrons  at U=0t. The   
spectrum in panel a) of Fig.\ref{4by4_lattice_spectrum_Ne14}  exhibits an increase in the 
density of energy levels, compared to the one at half-filling, pointing to
its very correlated nature.

The DOS 
${\cal{N}}(\omega)$ [Eq.(\ref{densityofstates})]
for  14 electrons 
at U=4t
 is shown in panel a) of Fig.\ref{densitystates_4by4_Ne14_U4}. The calculations have been
carried out by 
approximating the  ($N_{e}$ $\pm$ 1)-electron systems with
$n_{T}$=5   HF-transformations
along the lines described in Sec. \ref{spectral-H2D}.
A Lorentzian folding of width  $\Gamma$=0.2t has been used.
The  hole (blue) and particle (black) spectral functions 
are   
displayed in panel b) of the figure.
In 
this case, Eqs.(\ref{HW-holes-H}) and (\ref{HW-particles-H}) provide us with 
140 hole  and 180  particle solutions.  The chemical potential is now 
located around $\omega$-U/2=-1.2t. The comparison with panel b) of Fig.\ref{densitystates_halfilling_4by4_U4} reveals 
that the structure
of the hole states for $\omega$-U/2 $<$ -2t 
[i.e., those with linear momenta $(\pm \pi/2,0)$, $(0,\pm \pi/2)$ and $(0,0)$]
remains to a large extent  intact. On the
other hand, a large fraction of the particle spectral weight observed  at half-filling 
for 1t $<$ $\omega$-U/2 $<$ 3t is removed. This
depletion occurs in favor of new states  near
$\omega$-U/2=0. The spectral
decomposition of the DOS clearly shows that it is states around the
Fermi surface that suffer the most pronounced changes with respect to
half-filling. As a result, the 
particle-hole 
symmetry in the DOS, observed in Fig.\ref{densitystates_halfilling_4by4_U4}, is suppressed 
for this doped lattice and the original gap dissapears.

%%%%%%%%%%%%%%%%%%%%%%%%%%%%%%%%%%%%%%%%%%%%%%%%%%%%%%%%%%%%%%%%%%%%%%%%%%%%%%%%%%%%%%%%%%%%%%%%%%%
%
%  FIGURE 12 OF THE PAPER
%
%%%%%%%%%%%%%%%%%%%%%%%%%%%%%%%%%%%%%%%%%%%%%%%%%%%%%%%%%%%%%%%%%%%%%%%%%%%%%%%%%%%%%%%%%%%%%%%%%%%
\begin{figure*} 
\includegraphics[width=1.00\textwidth]{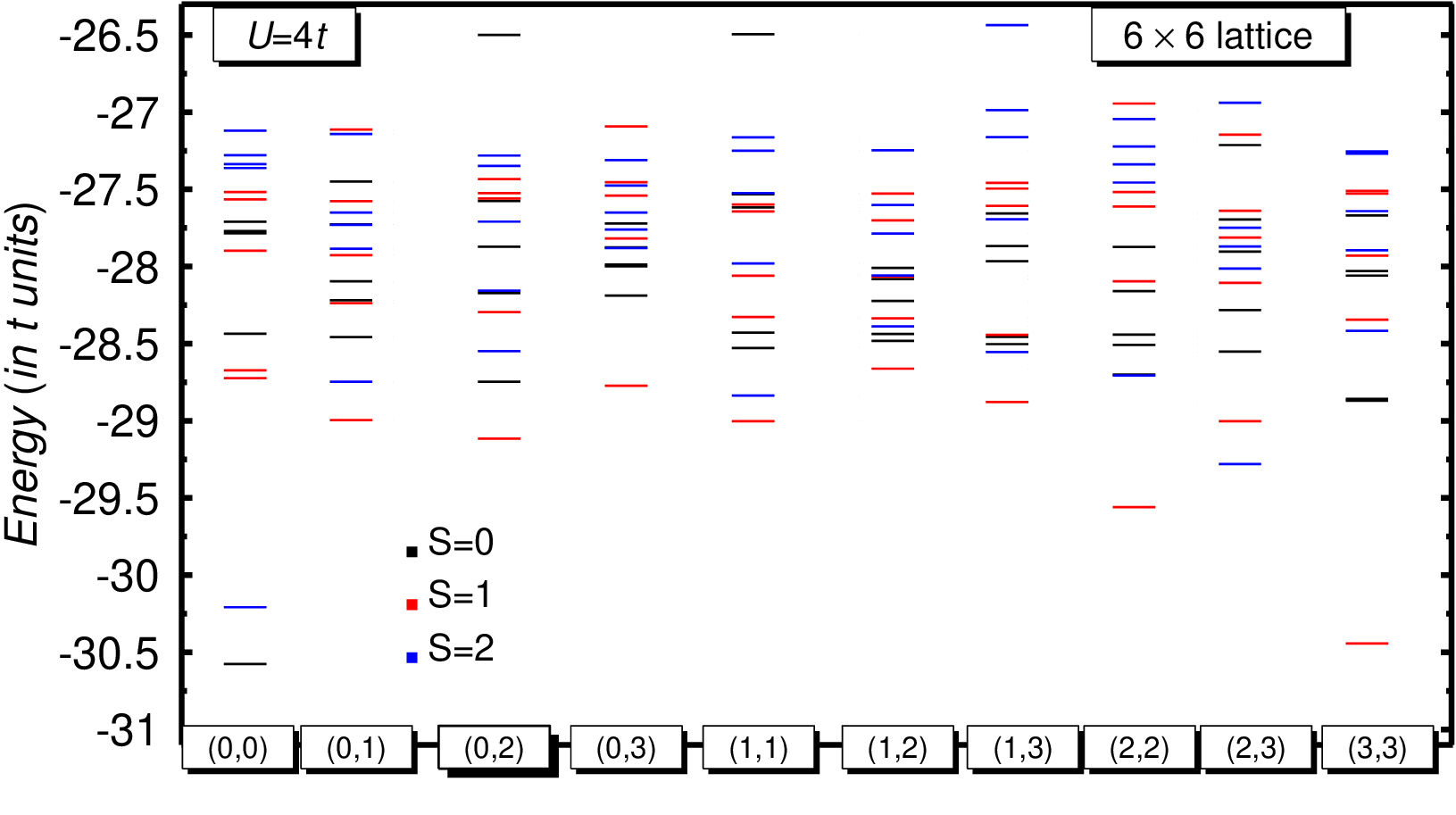} 
\caption{(Color online)  The same as Fig.\ref{4by4_lattice_spectrum_Ne16}
but for the half-filled $6 \times 6$ lattice at U=4t.
}
\label{6by6_lattice_spectrum_Ne36}
\end{figure*} 
%%%%%%%%%%%%%%%%%%%%%%%%%%%%%%%%%%%%%%%%%%%%%%%%%%%%%%%%%%%%%%%%%%%%%%%%%%%%%%%%%%%%%%%%%%%%%%%%%%%
%
%  END OF FIGURE 12 OF THE PAPER
%
%%%%%%%%%%%%%%%%%%%%%%%%%%%%%%%%%%%%%%%%%%%%%%%%%%%%%%%%%%%%%%%%%%%%%%%%%%%%%%%%%%%%%%%%%%%%%%%%%%%

In panel a) of Fig.\ref{densitystates_4by4_Ne14_U8}, we have plotted the DOS 
${\cal{N}}(\omega)$ [Eq.(\ref{densityofstates})]
for the  $4 \times 4$  lattice with  14 electrons 
at U=8t. The calculations were performed 
with $n_{T}$=5 HF-transformations and a folding $\Gamma$=0.2t has been used.
The  hole (blue) and particle (black) spectral functions 
are   
displayed in panel b) of the figure.
Our DOS and
spectral functions can
be compared with the ones, obtained using the Lanczos
method, shown in Figs.3 and 4  of Ref.57. The main qualitative features of the DOS are
well reproduced, namely the prominent peaks around 
$\omega$-U/2=-4t, -3t, -2t and 4t. As can be noted from panel b)
of Fig.\ref{densitystates_4by4_Ne14_U8}, the 
chemical potential is now located around $\omega$-U/2=-2.4t. 
One of the main features of the DOS is that it displays a
pronounced pseudogap which, as can be seen from 
Figs. \ref{densitystates_halfilling_4by4_U8} and 
\ref{densitystates_4by4_Ne14_U8}, results mainly from  pulling particle strenghts 
into the  (half-filling) gap  combined with  sizeable 
contributions of particle states around $\omega$-U/2=4t. 
We note, that the pseudogap problem in the doped 2D Hubbard 
model  has also received attention within the framework 
of quantum cluster approaches. \cite{maier2005}

We have also performed calculations for the $4 \times 4$
lattice
with 14 electrons at U=2t, 12t and 20t. From these 
calculations, and the results already discussed above for
the cases U=4t and 8t, we conclude that upon dopping
with two holes the original gap observed at half-filling
dissapears for U=2t and 4t, a pseudogap is developed 
around the noninteracting bandwidth W=8t while 
for the larger interaction strengths U=12t and U=20t
the gap is not filled. Similar conclusions 
have been obtained within the Lanczos framework.
\cite{Dagotto-RC-3184}

\subsection{The square $6 \times 6$ lattice} 
\label{6by6results}

Finally, let us turn our attention to the 
half-filled $6 \times 6$ lattice at U=4t. The dimensions 
$n_{RH}$($S=0,{\xi}_{x},{\xi}_{y}) \approx 2 \times 10^{17}$, $n_{RH}$($S=1,{\xi}_{x},{\xi}_{y}) \approx 6 \times 10^{17}$
and $n_{RH}$($S=2,{\xi}_{x},{\xi}_{y}) \approx 6 \times 10^{17}$ of the corresponding 
restricted Hilbert spaces are far too large for a brute force diagonalization to be feasible.
Other
approximate
methods are then called for, not only to describe ground state properties but also 
to access the  excitation spectrum in this relatively large lattice
for which information is rather scarce. In this case, the number of  variational parameters 
in our approximation
is $n_{var}$($S=0,{\xi}_{x},{\xi}_{y})$=2592, $n_{var}$($S=1,{\xi}_{x},{\xi}_{y})$=2596
and $n_{var}$($S=2,{\xi}_{x},{\xi}_{y})$=2600, respectively. Therefore, the half-filled $6 \times 6$ lattice 
represents a very  challenging testing ground for our symmetry-projected configuration mixing approximation.

In Fig.\ref{6by6_lattice_spectrum_Ne36}, we show  the energies
${\epsilon}_{\alpha}^{\Theta}$ obtained, via Eq.(\ref{diagonalizationfinal}), for the ten essentially different 
pairs of linear momentum quantum numbers (0,0), (0,1), (0,2), (0,3), (1,1), (1,2), (1,3), (2,2), (2,3) and 
(3,3). For 
each of them we have plotted 
the first five solutions with spins S=0,1, and 2. The ground state corresponds to the 
${\Theta}^{0}$=(0,0,0) configuration with energy 
${\epsilon}_{1}^{{\Theta}^{0}}$=-30.5766t. This can be compared with the energy obtained 
using state-of-the-art auxiliary-field MC, -30.89(1)t. \cite{Shiweiprivate}

%%%%%%%%%%%%%%%%%%%%%%%%%%%%%%%%%%%%%%%%%%%%%%%%%%%%%%%%%%%%%%%%%%%%%%%%%%%%%%%%%%%%%%%%%%%%%%%%%%%%%%%%%%%
%
%  FIGURE 13 OF THE PAPER
%
%%%%%%%%%%%%%%%%%%%%%%%%%%%%%%%%%%%%%%%%%%%%%%%%%%%%%%%%%%%%%%%%%%%%%%%%%%%%%%%%%%%%%%%%%%%%%%%%%%%%%%%%%%
\begin{figure*} 
\includegraphics[width=1.00\textwidth]{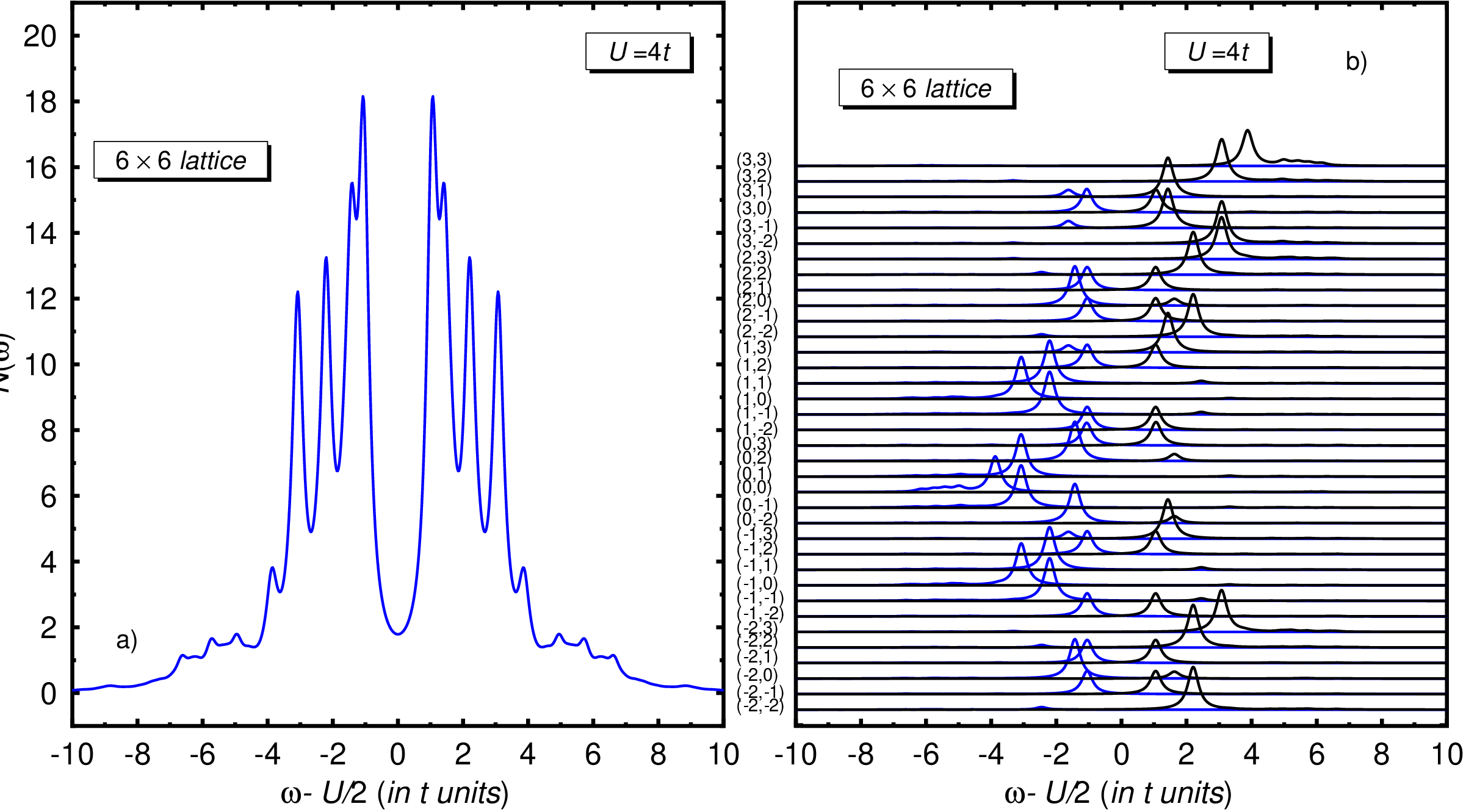} 
\caption{(Color online) The same as Fig.\ref{densitystates_halfilling_4by4_U4} 
but for the half-filled $6 \times 6$ lattice at U=4t.
}
\label{densitystates_halfilling_6by6_U4} 
\end{figure*}
%%%%%%%%%%%%%%%%%%%%%%%%%%%%%%%%%%%%%%%%%%%%%%%%%%%%%%%%%%%%%%%%%%%%%%%%%%%%%%%%%%%%%%%%%%%%%%%%%%%%%%%%%%%
%
%  END OF FIGURE 13 OF THE PAPER
%
%%%%%%%%%%%%%%%%%%%%%%%%%%%%%%%%%%%%%%%%%%%%%%%%%%%%%%%%%%%%%%%%%%%%%%%%%%%%%%%%%%%%%%%%%%%%%%%%%%%%%%%%%%

From Fig.\ref{6by6_lattice_spectrum_Ne36}, we realize that the first excited state 
corresponds to a 
 $\Theta=(1,3,3)$  configuration [with linear 
 momenta $(\pi,\pi)$]. The energy difference 0.1331t between this $(\pi,\pi)$-configuration and 
 the ground state is smaller than the corresponding value 0.1651t for the 
 half-filled $4 \times 4$ system. On the other hand, similar to the 
 half-filled $2 \times 4$ and $4 \times 4$ lattices, most of the 
 first excited states for each combination  $({\xi}_{x},{\xi}_{y})$ 
 have spin S=1, exception made of $(0,0)$ and $(2,3)$ for which 
 an S=2 quintet appears. Note that our calculations predict the lowest-lying 
 (1,0,1), (1,1,1) and (1,2,3) solutions to be quite close in energy. As with the other 
 half-filled lattices studied, we find only a handful of excited states within an energy
 window of t from the ground state.
 
The DOS 
${\cal{N}}(\omega)$ [Eq.(\ref{densityofstates})]
for the 
half-filled $6 \times 6$ lattice 
at U=4t
is shown in panel a) of Fig.\ref{densitystates_halfilling_6by6_U4}. The calculations 
have been carried out with $n_{T}$=5 HF-transformations 
along the lines described in Sec. \ref{spectral-H2D}. In 
this case, Eqs.(\ref{HW-holes-H}) and (\ref{HW-particles-H}) provide us with 
360 hole and particle solutions. A Lorentzian folding of width  $\Gamma$=0.2t has been used.
The DOS for this large lattice shows  a
clear Hubbard gap  $\Delta_H = U/2 = 2t$. The fact that the gap remains
intact in going, at half-filling, from the $4 \times 4$ to the $6 \times 6$ lattice is
consistent with previous studies within the DCA which show that it is preserved
even in the TDL.  \cite{moukouri2001, huscroft2001,aryanpour2003}
We note, that the DMFT (i.e., $N_{c}=1$) does not predict a gap for 
U=4t. It is only for a larger number $N_{c}$ of clusters  
that the gap
starts to develop in these studies at the TDL. \cite{maier2005}
Back to our DOS, we see again from the spectral
decomposition, shown in panel b) of Fig.\ref{densitystates_halfilling_6by6_U4}, that 
it is states at
 [i.e., $(\pm 2\pi/3,\pm \pi/3)$, $(\pm \pi/3,\pm 2\pi/3)$, $(\pi,0)$
and $(0,\pi)$] 
or close to the noninteracting Fermi surface
that contribute the
largest spectral weight to the prominent hole and particle peaks 
around $\omega$-U/2=-t and $\omega$-U/2=t. In general, the DOS
for this finite size
lattice is still highly peaked, though features should smooth
out as one approaches the TDL. Systematic studies  for the $8 \times 8$
and $10 \times 10$ lattices are in progress and will be presented
elsewhere.

\section{Conclusions}
\label{Conclusions-work}
How to accurately describe many-fermion systems with  approximate methods, which truncate the complete
expansion of the wave functions to a numerically feasible number of configurations, is a central 
question in nuclear structure theory, quantum chemistry, and condensed matter physics. To this end, in the 
present study we have explored an alternative avenue for the 2D Hubbard model. The 
main accomplishments of the present study are:

\begin{itemize}

\item We have presented a powerful methodology of a VAP configuration mixing scheme, originally devised for 
the nuclear many-body problem, but 
not yet used to study ground and excited states, with well 
defined quantum numbers, of the 2D Hubbard model with nearest-neighbor hopping and  PBC.

\end{itemize}

Our scheme relies on the Ritz variational principle to construct, throughout a chain 
of VAP calculations, a truncated basis consisting of a few (orthonormalized) symmetry-projected HF states. 
The simple structure of 
the projected wave functions employed, combined with a fast minimization algorithm, allows
 to keep low computational cost in building our basis.  
A further diagonalization of the Hamiltonian within such a basis 
allows to account, in a similar fashion, for residual correlations in the ground and excited states.

\begin{itemize}

\item Due to the simple structure of the wave functions in our approximation, we can construct an ansatz
[Eqs. (\ref{hole-wf-spectral} ) and (\ref{particle-wf-spectral})], whose 
flexibility is well-controlled by the number of HF-transformations included, to approximate the 
ground state of the $(N_{e} \pm 1)$-electron system. This allows us to 
determine one-electron affinities and ionization potentials  as well as to
access the spectral weight of states with different linear momentum quantum
numbers in the calculation of spectral functions and the corresponding density of states.

\end{itemize}

\begin{itemize}

\item We have shown that  our approximation gives accurate results, as compared with exact
energies,  for the   $2 \times 4$ and $4 \times 4$ lattices. We have also provided the 
low-lying spectrum of the $6 \times 6$ lattice which, to the best of our knowledge, has not been 
reported in the literature. Our ground state energy for this lattice compares well with results
from state-of-the-art auxiliary-field Monte Carlo calculations.

\end{itemize}

Regarding the physics of the 2D Hubbard model, we have discussed the trends, in going 
from the $2 \times4$ to the $4 \times 4$ and $6 \times 6$ half-filled lattices, of both 
the low-lying spectra and the spectral functions as well as the corresponding density of states.
We have found that the ground states correspond to configurations with spin zero and linear 
momenta (0,0). We have also found that most of the lowest-lying excited states display 
spin S=1. The doped systems with 14 and 15 electrons in the  $4 \times 4$ lattice have also
been considered. The ground states of such systems correspond to configurations with linear
momenta different from zero. 

Special attention has been paid
to the spectral weight of states with different linear momentum quantum
numbers. We have compared the DOS predicted within our approximation with
the one obtained using an exact diagonalization
for the half-filled $2 \times 4$ lattice
 and found an excellent 
agreement between the two. Our results for the half-filled 
$4 \times 4$ lattice, at different on-site repulsions, agree qualitatively well 
with the ones obtained using the Lanczos method. \cite{Dagotto-RC-3184} For all the considered 
half-filled lattices, a Hubbard gap is predicted within our 
approximation. In particular, the fact that this gap   
persists in going from the $4 \times 4$ to the $6 \times 6$ system is
consistent with previous studies within cluster extensions to dynamical
mean field theory which show that it is preserved
even in the thermodynamic limit. As opposed to
the half-filled case, the particle-hole symmetry in the DOS 
is removed when doping is  present 
in the system. From the calculations for 14 electrons 
in the $4 \times 4$ lattice we conclude that for on-site 
repulsions smaller than the noninteracting bandwidth 
the (half-filling) gap dissapears, a pseudogap develops
around U=8t while the gap is not filled for larger 
U values. These results agree well with similar conclusions 
extracted from Lanczos calculations. \cite{Dagotto-RC-3184}
We have also
found the remarkable result that all the lowest-lying  S=1/2 states
in the spectrum
of the $4 \times 4$ lattice with 15 electrons can be reasonably
well described by a wave function of  the form (\ref{hole-wf-spectral})
in which no orbital relaxation is accounted for. 

One important  feature of the scheme presented in this study is that it leaves 
ample space for further improvements and research. First, the 
number of symmetry-projected configurations in our basis set 
can be easily increased. Second, we 
could still incorporate particle number symmetry breaking and restoration in our 
configuration mixing scheme to access even more correlations. Our
 methods could  be useful even for more complicated 
lattices like the honeycomb one. An extension of the considered 2D t-U Hubbard Hamiltonian 
to the 
$t-t^{'}-t^{''}-U$  case is also straightforward, allowing the study 
of several interesting issues like indications 
of spin-charge separation  in 2D systems (see, for example, Ref. 65 and references therein). Last 
but not least, not only the configuration mixing scheme applied in the present work
but also the full hierarchy of approximations discussed in Ref. 43
can be implemented for the molecular Hamiltonian in the realm of quantum chemistry, within the 
already successful PQT.  \cite{PQT-reference-1,PQT-reference-2} 
Work along 
these avenues is in progress.

We would like to stress that the cost of the symmetry-projected calculations described 
in this work has the same scaling as mean-field methods. \cite{PQT-reference-1,PQT-reference-2} 
This statement is true as long as the number of grid points required in the symmetry restoration
remains relatively constant (this is usually the case for spin and number projection). Note, however, that
the restoration of translational symmetry in Hubbard lattices with PBC requires a number of grid
points equal to  $N_{sites}$. This makes the cost of our calculations ${\cal{O}}(N_{sites})$ more expensive 
than the Hartree-Fock method, which is still a very reasonable scaling. 
The computational effort 
is mainly concentrated in looping over grid points for the evaluation of matrix elements. This 
task is trivially parallelizable and one can thus easily reach clusters larger than the ones 
considered in this study. Preliminary calculations for the half-filled and doped $8 \times 8$ and 
$10 \times 10$ lattices are in progress.

A discussion of the limitations of our method is also in order here. Evidently, our method relies on the 
Hamiltonian having good symmetries. The lower the number of symmetries of a given Hamiltonian, the lower the  correlations
that can be accounted for by means of symmetry restoration. On the other hand, the most interesting quantum behavior 
is found in systems where symmetries are present. Second, it is known that
the correlation energy
per particle obtained with approaches based on a single symmetry-projected determinant \cite{PQT-reference-1,PQT-reference-2} 
decays as the lattice size
increases. We observe that the error in the   energy per site is larger  in the case of the half-filled $6 \times 6$ lattice than 
in the $4 \times 4$ one, although in both cases we have restricted the present study to $m$=5 transformations.
One can, however, increase the number of transformations to maintain  the quality of our wave functions. In principle, if 
the number of transformations 
is equal to the size of the restricted Hilbert subspace the method becomes exact. In practice, one can only
 hope that 
the number of transformations needed to access the relevant physics of the considered lattices is relatively low.

Finally, we believe that the finite size calculations discussed in the present work are complementary to 
other approaches  where impurity solvers play an important role. However, the symmetries to be broken 
and restored in the impurity-bath and bath Hamiltonians will depend on the details of the case considered. \cite{Zgid}

\begin{acknowledgments}

This work is supported by the National Science Foundation
under grants CHE-0807194 and CHE-1110884, and the Welch
Foundation (C-0036). The authors would like to thank both 
Prof. Shiwei Zhang and Dr. Eric Neuscamman for providing 
us with unpublished results. We also thank Dr.  Donghyung Lee
for providing us with exact diagonalization energy results for the $2 \times 4$
lattice.

\end{acknowledgments}

\appendix

\begin{widetext}
\section{Symmetry-projected matrix elements between two Slater determinants 
$ | {\cal{D}}^{i} \rangle$ and $ | {\cal{D}}^{k} \rangle$
}
\label{App-1}
In this appendix, we present 
the expressions for the matrix elements 
${\cal{H}}_{\Sigma {\Sigma}^{'}}^{i k \Theta}=
\langle {\cal{D}}^{i} | \hat{H}_{Hub} \hat{P}_{\Sigma {\Sigma}^{'}}^{\Theta}| {\cal{D}}^{k} \rangle$
and ${\cal{N}}_{\Sigma {\Sigma}^{'}}^{i k \Theta}=
\langle {\cal{D}}^{i} | \hat{P}_{\Sigma {\Sigma}^{'}}^{\Theta}| {\cal{D}}^{k} \rangle$
required 
to compute the kernels ${\cal{H}}_{\Sigma {\Sigma}^{'}}^{m \Theta}$
and ${\cal{N}}_{\Sigma {\Sigma}^{'}}^{m \Theta}$ in 
Eq.(\ref{ojo-ojo}) of Sec. \ref{formalism-H2D-excited}. Note that the matrix elements 
required in Eq.(\ref{energy-vampir})
of Sec. \ref{formalism-H2D} 
 are just a particular case  where both Slater determinants are the same. Here, and in what 
 follows, we keep our notation as close as possible to the one already used for the 1D Hubbard model. 
\cite{Carlos-Hubbard-1D} Both ${\cal{H}}_{\Sigma {\Sigma}^{'}}^{ik \Theta}$ and ${\cal{N}}_{\Sigma {\Sigma}^{'}}^{i k \Theta}$
read

\begin{eqnarray} \label{building-n}
{\cal{H}}_{\Sigma {\Sigma}^{'}}^{i k \Theta}
&=& 
\frac{2S+1}{8 {\pi}^{2} N_{sites}}
\sum_{{\bf{j}}}
e^{-i {\bf{k}}_{\xi} {\bf{j}}}
\int d\Omega D_{\Sigma {\Sigma}^{'}}^{S *}(\Omega)  
h^{ik}(\Omega,{\bf{j}})
n^{ik}(\Omega,{\bf{j}})
\nonumber\\
{\cal{N}}_{\Sigma {\Sigma}^{'}}^{i k \Theta}
&=&
\frac{2S+1}{8 {\pi}^{2} N_{sites}} 
\sum_{{\bf{j}}}
e^{-i \bf{k}_{\xi} \bf{j}}
\int d\Omega D_{\Sigma {\Sigma}^{'}}^{S *}(\Omega)  
n^{ik}(\Omega,{\bf{j}})
\end{eqnarray}
where,  ${\bf{k}}_{\xi}= \left(k_{{\xi}_{x}},k_{{\xi}_{y}}\right)
=
\left(\frac{2\pi {\xi}_{x}}{N_{x}},\frac{2\pi {\xi}_{y}}{N_{y}} \right)$ and 
${\bf{j}}=\left(j_{x},j_{y}\right)$, respectively.
For the  gauge-rotated 
norm 

\begin{eqnarray}
n^{ik}(\Omega,{\bf{j}}) = det_{N_{e}} {\cal{X}}^{ik}(\Omega,{\bf{j}})
\end{eqnarray}
the determinant has to be taken over the $N_{e} \times N_{e}$ dimensional
occupied part of the matrix

\begin{eqnarray} \label{keepX}
{\cal{X}}_{ab}^{ik}(\Omega,{\bf{j}}) 
 = \left( {\cal{D}}^{i T} {\cal{S}}(\Omega,{\bf{j}}) {\cal{D}}^{k *} \right)_{ab}
\end{eqnarray}
with 

\begin{eqnarray}
{\cal{S}}_{{\boldsymbol{\alpha}}  \sigma {\sigma}^{'}}(\Omega,{\bf{j}}) 
=
{\cal{D}}_{\sigma {\sigma}^{'}}^{1/2}(\Omega)
e^{i {\bf{k}}_{\alpha} {\bf{j}}}
\end{eqnarray}

The gauge-rotated 
Hamiltonian  takes the form

\begin{eqnarray}
h^{ik}(\Omega,{\bf{j}}) = 
\frac{1}{2}
t^{ik}(\Omega,{\bf{j}}) + 
\frac{1}{2}
Tr
\left(
{\Gamma}^{i k} (\Omega,{\bf{j}}) {\rho}^{k i} (\Omega,{\bf{j}})
\right)
\end{eqnarray}
with 

\begin{align} \label{eq-seenergies}
t^{ik}(\Omega,{\bf{j}}) &= 
\sum_{{\boldsymbol{\alpha}} {\sigma}} 
\epsilon({\bf{k}}_{\alpha})
{\rho}_{{\boldsymbol{\alpha}} {\sigma},{\boldsymbol{\alpha}} {\sigma} }^{k i} (\Omega,{\bf{j}})
\nonumber\\
\epsilon({\bf{k}}_{\alpha}) &= -2 t \left( cos  k_{{\alpha}_{x}}  + cos k_{{\alpha}_{y}} \right)
\nonumber\\ 
{\rho}_{{\boldsymbol{\gamma}} {\sigma}^{'},{\boldsymbol{\alpha}} \sigma }^{k i} (\Omega,{\bf{j}})
& =
 \sum_{{\sigma}^{"} h h^{'}} {\cal{S}}_{{\boldsymbol{\gamma}}  {\sigma}^{'}{\sigma}^{"} } (\Omega,{\bf{j}}) 
 {\cal{D}}_{{\boldsymbol{\gamma}} {\sigma}^{"},h}^{k *}
\Big[{\cal{X}}_{h h^{'}}^{ik} (\Omega,{\bf{j}}) \Big]^{-1}
{\cal{D}}_{{\boldsymbol{\alpha}} \sigma, h^{'}}^{i}
\nonumber\\
{\Gamma}_{{\boldsymbol{\alpha}}  \sigma, {\boldsymbol{\gamma}}  {\sigma}^{'}}^{i k} (\Omega,{\bf{j}}) 
&=
 {\delta}_{\sigma {\sigma}^{'}} 
 {\delta}_{{\alpha}_{x} {\gamma}_{x}} 
  {\delta}_{{\alpha}_{y} {\gamma}_{y}} 
\epsilon({\bf{k}}_{\alpha})
 + 
 \frac{U}{N_{sites}}
\sum_{{\boldsymbol{\beta}} {\boldsymbol{\delta}}}
 {\delta}_{{\alpha}_{x}+ {\beta}_{x} - {\gamma}_{x} - {\delta}_{x} }^{0, \pm N_{x}}
  {\delta}_{{\alpha}_{y}+ {\beta}_{y} - {\gamma}_{y} - {\delta}_{y} }^{0, \pm N_{y}}
\times
\nonumber\\
&\times  
\Big[
 {\delta}_{\sigma {\sigma}^{'}} {\rho}_{{\boldsymbol{\delta}} -\sigma,{\boldsymbol{\beta}} -\sigma }^{k i} (\Omega,{\bf{j}})
 - \left( 1-  {\delta}_{\sigma {\sigma}^{'}}  \right)  {\rho}_{{\boldsymbol{\delta}} \sigma,{\boldsymbol{\beta}} -\sigma }^{k i} (\Omega,{\bf{j}})
\Big]   
\end{align}
where the product of generalized Kronecker deltas 
in ${\Gamma}^{i k} (\Omega,{\bf{j}})$
results from the transformation of 
the on-site interaction term in Eq.(\ref{HAM-HUB-2D}) 
to the 
momentum representation. As a consequence of 
the PBC, ${\delta}_{{\alpha}_{i}+{\beta}_{i}-{\gamma}_{i}-{\delta}_{i}}^{0; \pm N_{i}}$
is one  if  ${\alpha}_{i}+{\beta}_{i}-{\gamma}_{i}-{\delta}_{i}$
is either 0 or 
$\pm N_{i}$ and zero else.

%%%%%%%%%%%%%%%%%%%%%%%%%%%%%%%%%%%%%%%%%%%%%%%%%%%%%%%%%%%%%%%%%
%%%%%%%%%%%%%%%%%%%%%%%%%%%%%%%%%%%%%%%%%%%%%%%%%%%%%%%%%%%%%%%%%
%%%%%%%%%%%%%%%%%%%%%%%%%%%%%%%%%%%%%%%%%%%%%%%%%%%%%%%%%%%%%%%%%
%%%%%%%%%%%%%%%%%%%%%%%%%%%%%%%%%%%%%%%%%%%%%%%%%%%%%%%%%%%%%%%%%
\section{Symmetry-projected particle-hole matrix elements between two Slater determinants 
$ | {\cal{D}}^{i} \rangle$ and $ | {\cal{D}}^{k} \rangle$
}
\label{App-2}
In this appendix, we present 
the expressions for the matrix elements 
${\cal{H}}_{\Sigma {\Sigma}^{'}}^{i k \Theta;ph}
=
\langle {\cal{D}}^{i} | \hat{H}_{Hub} \hat{P}_{\Sigma {\Sigma}^{'}}^{\Theta}
{\hat{b}}^{\dagger}_{p}({\cal{D}}^{k}) {\hat{b}}_{h}({\cal{D}}^{k})
| {\cal{D}}^{k} \rangle
$
and 
${\cal{N}}_{\Sigma {\Sigma}^{'}}^{i k \Theta;ph}
=
\langle {\cal{D}}^{i} |  \hat{P}_{\Sigma {\Sigma}^{'}}^{\Theta}
{\hat{b}}^{\dagger}_{p}({\cal{D}}^{k}) {\hat{b}}_{h}({\cal{D}}^{k})
| {\cal{D}}^{k} \rangle
$
required to 
compute the kernels 
${\cal{K}}_{\Sigma {\Sigma}^{'}}^{m \Theta; ph}$
and ${\cal{R}}_{\Sigma {\Sigma}^{'}}^{m \Theta; ph}$
defining the variational equations discussed in 
Sec. \ref{formalism-H2D-excited}. Note that the matrix elements 
required in Eq.(\ref{gradient}) of Sec. \ref{formalism-H2D} 
 are just a particular case  where both Slater determinants are the same.
We obtain

\begin{eqnarray} \label{building-n-ph}
{\cal{H}}_{\Sigma {\Sigma}^{'}}^{i k \Theta;ph}
&=&
\frac{2S+1}{8 {\pi}^{2} N_{sites}} 
\sum_{{\bf{j}}}
e^{-i {\bf{k}}_{\xi} {\bf{j}}}
\int d\Omega D_{\Sigma {\Sigma}^{'}}^{S *}(\Omega)  
n^{ik}(\Omega,{\bf{j}})
h_{ph}^{ik}(\Omega,{\bf{j}})
\nonumber\\
{\cal{N}}_{\Sigma {\Sigma}^{'}}^{i k \Theta;ph}
&=&
\frac{2S+1}{8 {\pi}^{2} N_{sites}} 
\sum_{{\bf{j}}}
e^{-i {\bf{k}}_{\xi} {\bf{j}}}
\int d\Omega D_{\Sigma {\Sigma}^{'}}^{S *}(\Omega)  
n^{ik}(\Omega,{\bf{j}})
n_{ph}^{ik}(\Omega,{\bf{j}})
\end{eqnarray}
where, as in appendix \ref{App-1},  ${\bf{k}}_{\xi}= \left(k_{{\xi}_{x}},k_{{\xi}_{y}}\right)
=
\left(\frac{2\pi {\xi}_{x}}{N_{x}},\frac{2\pi {\xi}_{y}}{N_{y}} \right)$ and 
${\bf{j}}=\left(j_{x},j_{y}\right)$, respectively. On the other hand,

\begin{eqnarray}
n_{ph}^{ik}(\Omega,{\bf{j}}) =
\sum_{h{'} \in  {\cal{D}}^{(i)}} 
\Big[{\cal{X}}_{hh^{'}}^{ik}(\Omega,{\bf{j}}) \Big]^{-1}
{\cal{X}}_{ h^{'}p}^{ik}(\Omega,{\bf{j}}) 
\end{eqnarray}
with the indices $h$ ($p$) running over all the 
occupied (unoccupied) states in  $| {\cal{D}}^{k} \rangle$. The inverse 
$\Big[{\cal{X}}_{hh^{'}}^{ik}(\Omega,{\bf{j}}) \Big]^{-1}$
is taken over the occupied part of the matrix (\ref{keepX}). Finally,

\begin{eqnarray}
h_{ph}^{ik}(\Omega,{\bf{j}}) &=&
n_{ph}^{ik}(\Omega,{\bf{j}}) h^{ik}(\Omega,{\bf{j}})
+
\Big [{\cal{Y}}^{ki}(\Omega,{\bf{j}}) 
{\Gamma}^{i k} (\Omega,{\bf{j}})
{\overline{\cal{W}}}^{ki}(\Omega,{\bf{j}})
\Big]_{hp}
\end{eqnarray}
with the functions  ${\cal{Y}}(\Omega,{\bf{j}})$
and ${\overline{\cal{W}}}^{ki}(\Omega,{\bf{j}})$)
defined, for all the occupied $h$ and unoccupied $p$ states  in $| {\cal{D}}^{k} \rangle$,  as

\begin{eqnarray} \label{refY}
{\cal{Y}}_{h,{\boldsymbol{\alpha}} \sigma}^{ki}(\Omega,{\bf{j}})  &=&
\sum_{h^{'}}
\Big[{\cal{X}}_{hh^{'}}^{ik}(\Omega,{\bf{j}} )\Big]^{-1}
{\cal{D}}_{{\boldsymbol{\alpha}} \sigma,h^{'}}^{i}
\nonumber\\
{\overline{\cal{W}}}_{{\boldsymbol{\gamma}} {\sigma}^{'},p}^{ki}(\Omega,{\bf{j}}) &=&
\sum_{{\boldsymbol{\delta}} {\sigma}^{"} {\sigma}^{'''}}
\Big[   
1-  {\rho}^{k i} (\Omega,{\bf{j}})
\Big]_{{\boldsymbol{\gamma}} {\sigma}^{'},{\boldsymbol{\delta}} {\sigma}^{"} }
{\cal{S}}_{{\boldsymbol{\delta}} {\sigma}^{"}{\sigma}^{'''}}(\Omega,{\bf{j}})
{\cal{D}}_{{\boldsymbol{\delta}} {\sigma}^{'''},p}^{k * }
\end{eqnarray}

%%%%%%%%%%%%%%%%%%%%%%%%%%%%%%%%%%%%%%%%%%%%%%%%%%%%%%%%%%%%%%%%
%%%%%%%%%%%%%%%%%%%%%%%%%%%%%%%%%%%%%%%%%%%%%%%%%%%%%%%%%%%%%%%%%
%%%%%%%%%%%%%%%%%%%%%%%%%%%%%%%%%%%%%%%%%%%%%%%%%%%%%%%%%%%%%%%%%
%%%%%%%%%%%%%%%%%%%%%%%%%%%%%%%%%%%%%%%%%%%%%%%%%%%%%%%%%%%%%%%%%
\section{Symmetry-projected  matrix elements between two Slater determinants 
$ | {\cal{D}}^{i} \rangle$ and $ | {\cal{D}}^{k} \rangle$ for spectral 
functions
}
\label{App-3}
In this appendix, we present the 
computation of the kernels ${\cal{H}}^{{\Theta}^{-}}$
and ${\cal{N}}^{{\Theta}^{-}}$ required in Eq.(\ref{HW-holes-H})
as well as of the kernels ${\cal{H}}^{{\Theta}^{+}}$
and ${\cal{N}}^{{\Theta}^{+}}$ in Eq.(\ref{HW-particles-H}). Both 
${\cal{H}}^{{\Theta}^{-}}$
and ${\cal{N}}^{{\Theta}^{-}}$ read

\begin{eqnarray}
{\cal{N}}_{i h \sigma; k h^{'} {\sigma}^{'}}^{{\Theta}^{-}} &=&
\frac{2}{8 {\pi}^{2} N_{sites}} 
\sum_{{\bf{j}}}
e^{-i {\bf{k}}_{\xi} {\bf{j}}}
\int d\Omega D_{\sigma {\sigma}^{'}}^{1/2 *}(\Omega)  
n^{ik} (\Omega,{\bf{j}}) n_{h h^{'}}^{ik} (\Omega,{\bf{j}})
\nonumber\\
{\cal{H}}_{i h \sigma; k h^{'} {\sigma}^{'}}^{{\Theta}^{-}}  &=&
\frac{2}{8 {\pi}^{2} N_{sites}} 
\sum_{{\bf{j}}}
e^{-i {\bf{k}}_{\xi} {\bf{j}}}
\int d\Omega D_{\sigma {\sigma}^{'}}^{1/2 *}(\Omega)  
n^{ik} (\Omega,{\bf{j}}) h_{h h^{'}}^{ik} (\Omega,{\bf{j}})
\end{eqnarray}
with the vector 
${\bf{k}}_{\xi}= \left(k_{{\xi}_{x}^{-}},k_{{\xi}_{y}^{-}}\right)
=
\left(\frac{2\pi {\xi}_{x}^{-}}{N_{x}},\frac{2\pi {\xi}_{y}^{-}}{N_{y}} \right)$
while i,k = 1, $\dots$ $n_{T}$, $h,h^{'}= 1, \dots N_{e}$ and $\sigma, {\sigma}^{'}=\pm 1/2$. On the other hand

\begin{eqnarray}
 n_{h h^{'}}^{ik} (\Omega,{\bf{j}}) = \Big[{\cal{X}}_{h^{'} h}^{ik} (\Omega,{\bf{j}}) \Big]^{-1}
\end{eqnarray}
and

\begin{eqnarray}
 h_{h h^{'}}^{ik} (\Omega,{\bf{j}})  = \Big[{\cal{X}}_{h^{'} h}^{ik} (\Omega,{\bf{j}}) \Big]^{-1}
h^{ik} (\Omega,{\bf{j}}) 
-
\Big[{\cal{Y}}^{k i} (\Omega,{\bf{j}})
{\Gamma}^{i k} (\Omega,{\bf{j}})
{\cal{Z}}^{k i} (\Omega,{\bf{j}})
\Big]_{h^{'}h}
\end{eqnarray}
respectively. The function ${\cal{Z}}_{{\boldsymbol{\gamma}} {\sigma}^{'},h}^{k i} (\Omega,{\bf{j}})$  reads

\begin{eqnarray}
{\cal{Z}}_{{\boldsymbol{\gamma}} {\sigma}^{'},h}^{k i} (\Omega,{\bf{j}}) =
\sum_{h^{"} \sigma^{"}} 
S_{{\boldsymbol{\gamma}} {\sigma}^{'}  {\sigma}^{"}} (\Omega,{\bf{j}})
D_{{\boldsymbol{\gamma}} {\sigma}^{"},h^{"}}^{k *}
\Big[{\cal{X}}_{h^{"} h}^{ik} (\Omega,{\bf{j}}) \Big]^{-1}
\end{eqnarray}
while ${\cal{Y}}_{h^{'}, {\boldsymbol{\alpha}} \sigma}^{k i} (\Omega,{\bf{j}})$ is given in
Eq.(\ref{refY}). 

The norm  
and Hamiltonian
overlaps  in Eq.(\ref{HW-particles-H}) read

\begin{eqnarray}
{\cal{N}}_{i p \sigma, k p^{'} {\sigma}^{'}}^{{\Theta}_{+}} &=&
\frac{2}{8 {\pi}^{2} N_{sites}} 
\sum_{{\bf{j}}}
e^{-i {\bf{k}}_{\xi} {\bf{j}}}
\int d\Omega D_{\sigma {\sigma}^{'}}^{1/2 *}(\Omega)  
n^{ik} (\Omega,{\bf{j}}) n_{p p^{'}}^{ik} (\Omega,{\bf{j}})
\nonumber\\
{\cal{H}}_{i p \sigma, k p^{'} {\sigma}^{'}}^{{\Theta}_{+}} &=&
\frac{2}{8 {\pi}^{2} N_{sites}} 
\sum_{{\bf{j}}}
e^{-i {\bf{k}}_{\xi} {\bf{j}}}
\int d\Omega D_{\sigma {\sigma}^{'}}^{1/2 *}(\Omega)  
n^{ik} (\Omega,{\bf{j}}) h_{p p^{'}}^{ik} (\Omega,{\bf{j}})
\end{eqnarray}
with  the vector 
${\bf{k}}_{\xi}= \left(k_{{\xi}_{x}^{+}},k_{{\xi}_{y}^{+}}\right)
=
\left(\frac{2\pi {\xi}_{x}^{+}}{N_{x}},\frac{2\pi {\xi}_{y}^{+}}{N_{y}} \right)$
while i,k = 1, $\dots$ $n_{T}$, $p,p^{'}= N_{e}+1, \dots  2N_{sites}$ and $\sigma, {\sigma}^{'}=\pm 1/2$. On the other hand,

\begin{eqnarray}
 n_{p p^{'}}^{ik} (\Omega,{\bf{j}}) = {\cal{X}}_{p p^{'}}^{ik} (\Omega,{\bf{j}})
- \sum_{h h^{'}} {\cal{X}}_{p h}^{ik} (\Omega,{\bf{j}}) 
\Big[{\cal{X}}_{h h^{'}}^{ik} (\Omega,{\bf{j}}) \Big]^{-1}
{\cal{X}}_{h^{'} p^{'}}^{ik} (\Omega,{\bf{j}})
\end{eqnarray}
and

\begin{eqnarray}
 h_{p p^{'}}^{ik} (\Omega,{\bf{j}}) &=& n_{p p^{'}}^{ik} (\Omega,{\bf{j}}) h^{ik}(\Omega,{\bf{j}})
+
\Big[
{\cal{W}}^{i k} (\Omega,{\bf{j}}) {\Gamma}^{i k} (\Omega,{\bf{j}}) 
{\overline{\cal{W}}}^{k i} (\Omega,{\bf{j}})
\Big]_{p p^{'}}
\end{eqnarray}
respectively. The function ${\cal{W}}_{p, {\boldsymbol{\alpha}} \sigma}^{i k} (\Omega,{\bf{j}})$  is given by

\begin{eqnarray}
{\cal{W}}_{p, {\boldsymbol{\alpha}} \sigma}^{i k} (\Omega,{\bf{j}}) =
\sum_{{\boldsymbol{\beta}} {\sigma}^{'}}
{\cal{D}}^{i}_{{\boldsymbol{\beta}} {\sigma}^{'},p}
\Big[ 
1-{\overline{\rho}}^{ki}(\Omega,{\bf{j}})
\Big]_{{\boldsymbol{\beta}} {\sigma}^{'},{\boldsymbol{\alpha}} \sigma}
\end{eqnarray}
while ${\overline{\cal{W}}}_{{\boldsymbol{\gamma}} {\sigma}^{'},p^{'}}^{k i} (\Omega,{\bf{j}})$
is defined in Eq.(\ref{refY}).
\end{widetext}

\end{document}